\PassOptionsToPackage{sort&compress,numbers,super,square}{natbib}
\documentclass[journal=apcach,manuscript=article]{achemso}

\usepackage{microtype}
\usepackage{amsmath,amssymb,amsthm}
\usepackage{mathtools}
\usepackage{booktabs}
\usepackage{array}
\usepackage{enumitem}
\usepackage{xcolor}
\usepackage{tikz}
\usepackage{pgfplots}
\pgfplotsset{compat=1.17}
\usetikzlibrary{arrows.meta,positioning,shadows}
\usepackage[section]{placeins}

\AtBeginDocument{%
  \setlength{\paperwidth}{210mm}%
  \setlength{\paperheight}{297mm}%
  \pdfpagewidth=\paperwidth
  \pdfpageheight=\paperheight
  \newgeometry{
    top    = 6mm,
    bottom = 8mm,
    left   = 10mm,
    right  = 10mm
  }%
}

\theoremstyle{definition}

\newcommand{\phig}{\varphi}
\newcommand{\Jcost}{J}
\newcommand{\Jchem}{\Jcost_{\!\mathrm{chem}}}
\newcommand{\Zp}{Z}

\newcommand{\IEone}{\mathrm{IE}_1}
\newcommand{\EA}{\mathrm{EA}}
\newcommand{\code}[1]{{\footnotesize\path{#1}}}
\definecolor{modBlue}{RGB}{0,0,170}
\definecolor{issueRed}{RGB}{170,0,0}
\definecolor{fixGreen}{RGB}{0,130,0}
\definecolor{issueMagenta}{RGB}{180,0,150}

\newcommand{\edit}[1]{#1}

\title{A Noble-Gas-Centered Coordinate for Within-Period Atomic Property Trends}

\author{Jonathan Washburn}
\affiliation{Recognition Science; Recognition Physics Institute, Austin, Texas, USA}

\author{Megan Simons}
\affiliation{Recognition Science; Recognition Physics Institute, Austin, Texas, USA}

\author{Elshad Allahyarov}
\email{elshad.allakhyarov@case.edu}
\affiliation{Recognition Science; Recognition Physics Institute, Austin, Texas, USA}
\alsoaffiliation{Department of Physics, Case Western Reserve University, Cleveland, Ohio, USA}
\alsoaffiliation{Institut f\"ur Theoretische Physik II: Weiche Materie, Heinrich-Heine Universit\"at D\"usseldorf, Germany}
\alsoaffiliation{Theoretical Department, Joint Institute for High Temperatures, RAS, Moscow, Russia}


\SectionNumbersOn

\begin{document}

\begin{abstract}
We introduce a single dimensionless landscape function
$\Jchem(\rho) = \cosh(\rho \ln \phig) - 1$,
$\phig = (1+\sqrt{5})/2$, on the noble-gas-centred coordinate
$\rho = d/L_p \in [0,1)$, and show that it organizes four central
atomic observables- first ionization energy $\IEone$, electron
affinity $\EA$, Mulliken electronegativity $\chi_M$, and Pearson
chemical hardness $\eta$, on one periodic-table axis. The
outward step $\Delta\Jchem^{+}$ delivers $\IEone$; the inward gap
$\Delta\Jchem^{-} = \Jchem(1) - \Jchem(\rho)$ delivers $\EA$ and
$\eta$; $\chi_M$ follows by Mulliken's identity.

Three results establish the empirical content. (i) The
within-period $\IEone$ envelope reproduces the full
noble-gas-to-alkali ordering across periods 2--6: of 34 atoms
compiled across periods 2--4, 26 lie on the predicted monotone
descent and the 8 upward deviations occur exactly at the textbook
anomaly sites $\{p^3, d^5, f^7, s^2, d^{10}\}$. (ii) Two
golden-ratio identities,
$\IEone(G_p)/\IEone(G_{p+1}) \approx \phig^{1/4}$ on three heavy
noble-gas pairs and
$\IEone(\text{halogen})/\IEone(\text{alkali}) \approx \phig^2$ on
four within-period pairs, agree with NIST data to MAD
$\approx 1\%$ and $\approx 5\%$, respectively. (iii) The shared
kernel $\Delta\Jchem^{-}$ provides single-parameter analytical
fits to $\EA$ across periods 4--6 (MAE $0.3$--$0.4$~eV), to
Pearson hardness $\eta$ across periods 2--4 (MAE $\sim 1$~eV on
noble-gas maxima up to $10.8$~eV), and to Mulliken $\chi_M$ across
a 15-atom four-class benchmark ($R^2 = 0.73$). \edit{At the
period-averaged scale level, the shared-kernel relation
$\EA/\eta \approx C^{(p)}_{\mathrm{EA}}/C^{(p)}_{\eta}$ is
supported on period-4 NIST data: the empirical nine-atom mean
$\overline{\EA/\eta} = 0.180$ agrees with the predicted constant
$0.182$ to better than $1\%$, although individual-atom scatter
($\sigma \approx 0.13$) is much larger.} The framework thus assembles four
independent periodic-table observables under a single
golden-ratio cosh coordinate, providing a compact analytical
reference against which relativistic and shell-anomaly
corrections can be quantified.
\end{abstract}


\section{Introduction}
\label{sec:intro}

\edit{Within each period of the periodic table~\cite{vanspronsen1969,gordin2004,kaji2003,seaborg1996},
first ionization energy decreases monotonically from the
period-closing noble gas to the alkali metal, with upward bumps at
half-filled and completed-subshell sites $\{p^3, d^5, f^7, s^2,
d^{10}\}$. Electron affinity and Pearson chemical hardness show the
same broad descent, with related anomaly structure. Mulliken
electronegativity inherits the same broad trend because it is half
their sum.}

\edit{This paper asks whether two landscape steps derived from a
single dimensionless function $\Jchem(\rho)$ on the
noble-gas-centred coordinate $\rho = d/L_p \in [0, 1)$- 
the outward step $\Delta\Jchem^{+}$ and the gap
$\Delta\Jchem^{-}=\Jchem(1)-\Jchem(\rho)$,  can organize
$\IEone$, $\EA$, and (via Mulliken's identity) $\chi_M$, with
$\eta$ assigned the same kernel as $\EA$. The function is the
cosh form $\Jchem(\rho) = \cosh(\rho \ln \phig) - 1$, with
$\phig = (1+\sqrt{5})/2$, established in
Refs.~\cite{axioms_paper,dalembert_paper} as the unique reciprocal
cost on positive ratios under standard smoothness conditions
(Appendix~\ref{app:jcost}).\edit{\footnote{The uniqueness
statement and the $\phig$-power identities used below are also
formalized in Lean~4 in the Recognition Science library; see the
Data Availability statement.}} The present
paper takes $\Jchem$ as a fixed analytical input and develops
its empirical content on the periodic table.}

\edit{The four observables are conventionally treated separately:
$\IEone$ and $\EA$ from atomic spectroscopy and electronic-structure
calculations~\cite{NISTASD,koopmans,atkins}; electronegativity from
empirical scales~\cite{pauling,mulliken1934,allredrochow1958,allen1989,jensen1996,franco2019};
chemical hardness from derivative-based density-functional
arguments~\cite{parr1978,parrpearson1983,pearson,parryang1989,sen1991}.
For context, recent reviews on periodic-table form, shell
filling, and relativistic extensions to superheavy atoms include
Refs.~\cite{scerri2020,pyykko2012,pershina2019};
the present work combines IE$_1$, EA, electronegativity, and
chemical hardness in a single coordinate-based analysis.}

\edit{The empirical content reduces to three $\IEone$ predictions
and three derived checks. The $\IEone$ predictions are the heavy
noble-gas ratio $\IEone(G_p)/\IEone(G_{p+1}) \approx \phig^{1/4}$,
the halogen-to-alkali ratio $\phig^2$, and the within-period
monotone $\IEone$ envelope with the standard anomaly sites flagged
(Prediction~3, where periods 2--4 give 26 non-anomalous points on
the monotone trend and 8 deviations at the textbook anomaly
positions). The derived checks are a one-parameter EA fit across
periods 4--6, a one-parameter Pearson-hardness fit across periods
2--4, and a single-parameter Mulliken-$\chi_M$ fit on 15 atoms
across four chemical classes ($R^2 = 0.73$ globally,
Section~\ref{sec:data}).}

\edit{One closed-form expression
$\Delta\Jchem^{-}(\Zp) = \Jchem(1) - \Jchem(d(\Zp)/L_p)$ is used
for two observables: the EA proxy of Section~\ref{sec:ea} and the
hardness predictor $\kappa_{\mathrm{RS}}$ of
Section~\ref{sec:hardness} (the functional-equation framework of
Refs.~\cite{axioms_paper,dalembert_paper,ledger_gravity_paper,cocycle_tilings_paper}).
The shared-kernel form unifies two observables under one
analytical landscape gap with two independent per-period scales
fitted against the EA and $\eta$ data; the resulting ratio
benchmark $\EA/\eta \approx C^{(p)}_{\mathrm{EA}}/C^{(p)}_{\eta}$
is confirmed by the period-4 NIST data with empirical mean
matching the predicted constant to better than $1\%$
(Section~\ref{sec:shared_kernel_test}).}

This paper proposes a compact phenomenological,
coordinate-based framework that organizes broad within-period
trends in $\IEone$, $\EA$, $\chi_M$, and $\eta$ on a single
analytical axis, and provides a useful baseline against which
shell and relativistic corrections can be measured. The work is
\emph{not} an ab initio derivation of atomic structure, does
\emph{not} replace electronic-structure theory, does \emph{not}
derive shell anomalies (which we identify post hoc with textbook
sites), and does \emph{not} provide a universal predictor without
exclusions. Two of the headline ratio identities
($\phig^{1/4}$ for heavy noble-gas IE, $\phig^{2}$ for
halogen/alkali IE) are empirical regularities consistent with the
framework once a per-period absolute-energy scale $E_p$ is
admitted from data; a first-principles derivation of $E_p$ is
left to future work. ``Prediction'' in the rest of the paper
refers to a model-derived ordering or proxy expression evaluated
on a stated benchmark subset, not to a parameter-free quantitative
forecast.

\edit{The paper is organized as follows.
Section~\ref{sec:framework} fixes the notation.
Section~\ref{sec:landscape} constructs the landscape.
Section~\ref{sec:observables} assigns the four observables to
landscape steps. Section~\ref{sec:data} presents the three
$\IEone$ predictions and the EA, electronegativity, and hardness
checks against NIST and Pearson data, and
tests the shared-kernel ratio benchmark
(Section~\ref{sec:shared_kernel_test}).
Section~\ref{sec:conclusion} states the conclusions.}

\section{Mathematical Setup}
\label{sec:framework}
 The symbols used throughout the paper are given in Table~\ref{tab:nomenclature}.

  \begin{table}[ht]
\centering
\caption{Nomenclature used in the paper.}
\label{tab:nomenclature}
\small
\begin{tabular}{>{\raggedright\arraybackslash}p{2.4cm}>{\raggedright\arraybackslash}p{10.0cm}}
\toprule
Symbol & Meaning \\
\midrule
$Z$ & Atomic number of the element under discussion. \\
$G_p$ & Atomic number of the noble gas at the end of row $p$ of the periodic table. \\
$L_p$ & Length of row $p$, i.e.\ the number of elements in that row. \\
$d(Z)$ & Number of electrons needed for element $Z$ to reach the noble gas at the end of its row. \\
$v(Z)$ & Position of the element within its row, counted from the alkali-metal side. \\
$\rho(Z)$ & Normalized position in the row, defined as $d(Z)/L_p$. \\
$\Jchem(\rho)$ & Dimensionless landscape function used throughout the paper. \\
$\Delta\Jchem^{+}$ & Outward one-electron step: one electron farther from the noble-gas end of the row. \\
$\Delta\Jchem^{-}$ & \edit{Landscape gap to the fixed reference $\Jchem(1)$: $\Jchem(1)-\Jchem(d/L_p)$. Used as both the EA proxy and the hardness predictor $\kappa_{\mathrm{RS}}$.} \\
$\chi_{\mathrm{struct}}$ & \edit{Half-sum of the outward IE-step $\Delta\Jchem^{+}$ and the EA-side gap $\Delta\Jchem^{-}$.} \\
$\chi_M$ & Mulliken electronegativity, $\tfrac{1}{2}(\IEone+\EA)$. \\
$\eta$ & Pearson hardness, $\tfrac{1}{2}(\IEone-\EA)$. \\
$E_p$ & Period-dependent energy scale needed to convert the dimensionless landscape into eV. \\
\bottomrule
\end{tabular}
\end{table}

\edit{The model adopts three inputs: (i) the reciprocal cost
functional $\Jcost$, derived in
Refs.~\cite{axioms_paper,dalembert_paper,ledger_gravity_paper,cocycle_tilings_paper}
and summarized in Appendix~\ref{app:jcost}; (ii) the choice
$\phig = (1 + \sqrt{5})/2$ as the argument-rescaling factor, so
that $\Jcost$ is evaluated at $x = \phig^{\rho}$ for
$\rho \in [0, 1)$; and (iii) the standard period lengths $L_p$ and
noble-gas atomic numbers $G_p$ from the periodic table. The
discrete distance to the next noble gas is constructed below.}

\subsection{Discrete index $\delta(\Zp)$ and continuous coordinate distance $\rho(\Zp)$}
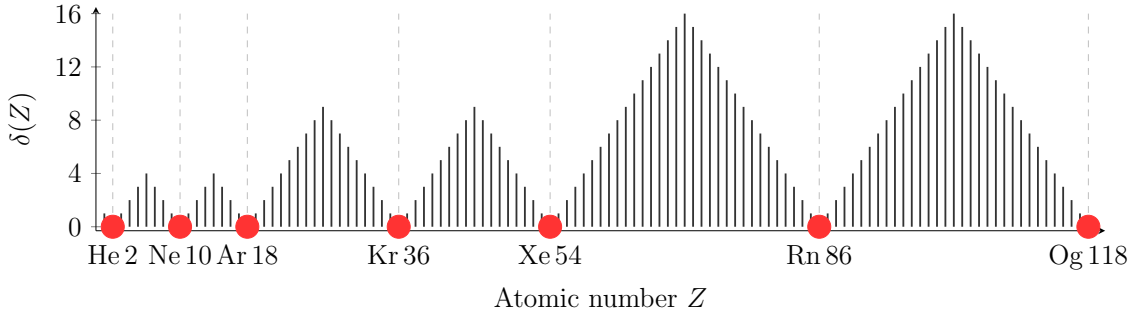
\begin{figure}[H]
\centering
\resizebox{0.768\textwidth}{!}{%
\begin{tikzpicture}
\begin{axis}[
  width=17cm, height=5cm,
  xlabel={Atomic number $Z$},
  ylabel={$\delta(Z)$},
  xmin=0, xmax=120,
  ymin=-0.3, ymax=16.5,
  xtick={2,10,18,36,54,86,118},
  xticklabels={He\,2, Ne\,10, Ar\,18, Kr\,36, Xe\,54, Rn\,86, Og\,118},
  ytick={0,4,8,12,16},
  axis lines=left,
  clip=false,
]
\addplot[gray!50, dashed, thin, forget plot] coordinates {(2,0)(2,16)};
\addplot[gray!50, dashed, thin, forget plot] coordinates {(10,0)(10,16)};
\addplot[gray!50, dashed, thin, forget plot] coordinates {(18,0)(18,16)};
\addplot[gray!50, dashed, thin, forget plot] coordinates {(36,0)(36,16)};
\addplot[gray!50, dashed, thin, forget plot] coordinates {(54,0)(54,16)};
\addplot[gray!50, dashed, thin, forget plot] coordinates {(86,0)(86,16)};
\addplot[gray!50, dashed, thin, forget plot] coordinates {(118,0)(118,16)};
\addplot[black!80, ycomb, thick] coordinates {
  (1,1)(2,0)(3,1)(4,2)(5,3)(6,4)(7,3)(8,2)(9,1)(10,0)
  (11,1)(12,2)(13,3)(14,4)(15,3)(16,2)(17,1)(18,0)
  (19,1)(20,2)(21,3)(22,4)(23,5)(24,6)(25,7)(26,8)(27,9)
  (28,8)(29,7)(30,6)(31,5)(32,4)(33,3)(34,2)(35,1)(36,0)
  (37,1)(38,2)(39,3)(40,4)(41,5)(42,6)(43,7)(44,8)(45,9)
  (46,8)(47,7)(48,6)(49,5)(50,4)(51,3)(52,2)(53,1)(54,0)
  (55,1)(56,2)(57,3)(58,4)(59,5)(60,6)(61,7)(62,8)(63,9)(64,10)
  (65,11)(66,12)(67,13)(68,14)(69,15)(70,16)(71,15)(72,14)(73,13)(74,12)
  (75,11)(76,10)(77,9)(78,8)(79,7)(80,6)(81,5)(82,4)(83,3)(84,2)(85,1)(86,0)
  (87,1)(88,2)(89,3)(90,4)(91,5)(92,6)(93,7)(94,8)(95,9)(96,10)
  (97,11)(98,12)(99,13)(100,14)(101,15)(102,16)(103,15)(104,14)(105,13)(106,12)
  (107,11)(108,10)(109,9)(110,8)(111,7)(112,6)(113,5)(114,4)(115,3)(116,2)(117,1)(118,0)
};
\addplot[only marks, mark=*, mark size=5pt, red!80, forget plot]
  coordinates {(2,0)(10,0)(18,0)(36,0)(54,0)(86,0)(118,0)};
\end{axis}
\end{tikzpicture}%
}%
\caption{Distance index $\delta(Z)$ from
  Eq.~\eqref{eq:defect} for all elements
$Z \leq 118$ (black bars). Noble gases (red dots) are the unique minima
$\delta = 0$.}
\label{fig:defect_profile}
\end{figure}

We attach to each element $\Zp \leq 118$ a discrete distance
$\delta(\Zp)$ from the filled-shell noble-gas state at the end of
its row $p$,
\begin{equation}
\delta(\Zp) = \min\bigl(v(\Zp),\; d(\Zp)\bigr),
\label{eq:defect}
\end{equation}
where $v(\Zp) = \Zp - G_{p-1}$ is the valence electron count
(electrons beyond the previous noble gas core) and
$d(\Zp) = G_p - \Zp$ is the distance to the next noble gas at the
end of the row\edit{, with
$G_p \in \{0,2,10,18,36,54,86,118\}$ the noble-gas atomic numbers
through Og}. By construction $\delta(\Zp) = 0$ exactly at the
noble gases, as seen in Figure ~\ref{fig:defect_profile}.

\edit{The symmetric index $\delta(\Zp) = \min(v(\Zp), d(\Zp))$ is
introduced only to make the noble-gas-centred structure of the
periodic table visually explicit
(Figure~\ref{fig:defect_profile}); the rest of the analysis uses
the asymmetric coordinate $\rho(\Zp) = d(\Zp)/L_p \in [0,1)$
defined in Eq.~\eqref{eq:rho}, with $\rho = 0$ at the
period-closing noble gas.}

\section{The Inverse Landscape}
\label{sec:landscape}

\edit{The discrete index $\delta(\Zp)$ of Section~\ref{sec:framework}
is converted to the continuous coordinate}
\begin{equation}
\rho(\Zp) \;=\; \frac{d(\Zp)}{L_p}
\;=\; \frac{G_p - \Zp}{G_p - G_{p-1}} \;\in\; [0, 1),
\label{eq:rho}
\end{equation}
\edit{with $L_p = G_p - G_{p-1}$ the period length, $\rho = 0$ at
the period-closing noble gas, and $\rho = (L_p - 1)/L_p < 1$ at the
alkali metal.}

\edit{The continuous chemical landscape $\Jchem$ is obtained by
evaluating the reciprocal cost functional $\Jcost$ of
Appendix~\ref{app:jcost} at the geometric scale
$x = \phig^{\,\rho(\Zp)}$, with the golden ratio
$\phig = (1+\sqrt{5})/2$ entering as the natural rescaling factor
on the noble-gas-centred coordinate (Appendix~\ref{app:jcost}).}
\begin{equation}
\Jchem(\Zp) \;=\; \Jchem(\rho) \;=\; \Jcost\!\bigl(\phig^{\,\rho(\Zp)}\bigr)
\;=\; \cosh\!\bigl(\rho(\Zp)\ln\phig\bigr) - 1.
\label{eq:landscape}
\end{equation}
\edit{Near $\rho = 0$,
$\Jchem(\rho) \approx \tfrac{1}{2}(\ln\phig)^2 \rho^2 \approx 0.116\,\rho^2$;
this small-step expansion is used in the noble-gas ratio derivation
of Section~\ref{sec:data}.}

\edit{$\Jchem$ is dimensionless, and we develop the framework
through two complementary routes: ratio-level identities (where
the period-dependent scale $E_p$ cancels exactly, giving
parameter-free predictions) and single-parameter fits (where $E_p$
is absorbed into one fitted scale per period, giving closed-form
analytical descriptions of the observables). Relativistic
corrections, many-electron correlation, orbital relaxation, and
the textbook subshell anomalies enter as quantified, well-localized
corrections on top of the smooth landscape.}

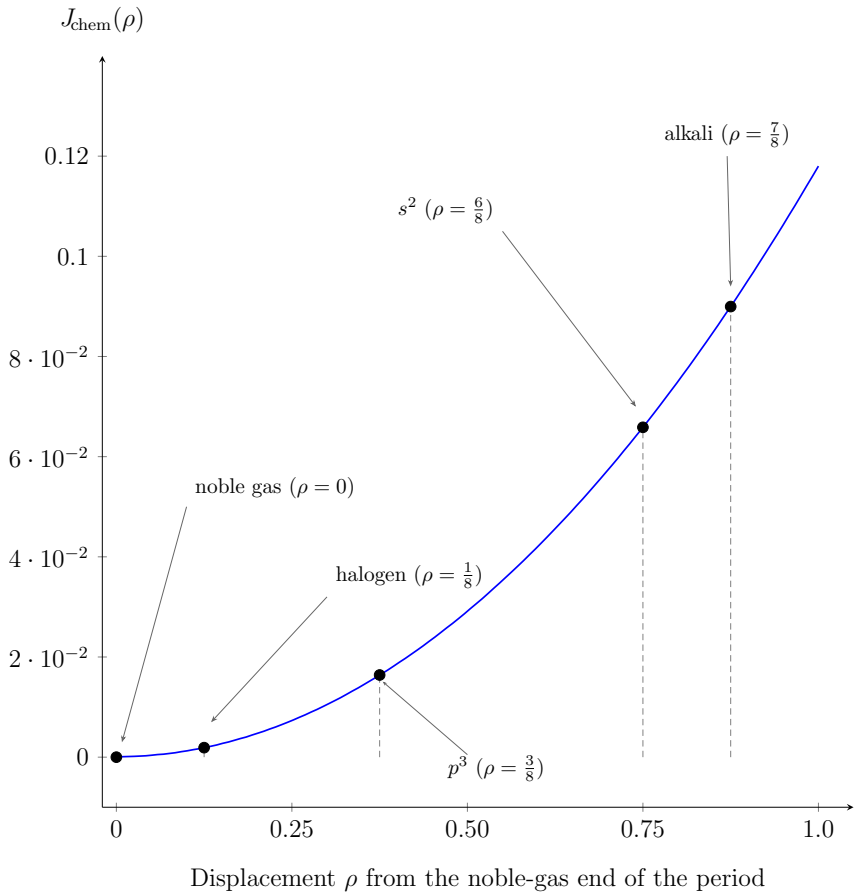
\begin{figure}[H]
\centering
\resizebox{0.576\textwidth}{!}{%
\begin{tikzpicture}
\begin{axis}[
  width=14cm,
  height=14cm,
  xlabel={Displacement $\rho$ from the noble-gas end of the period},
  ylabel={$\Jchem(\rho)$},
  xmin=-0.02, xmax=1.05,
  ymin=-0.010, ymax=0.140,
  xtick={0, 0.25, 0.5, 0.75, 1.0},
  xticklabels={$0$, $0.25$, $0.50$, $0.75$, $1.0$},
  ytick={0, 0.02, 0.04, 0.06, 0.08, 0.10, 0.12},
  axis lines=left,
  xlabel style={at={(axis description cs:0.5,-0.07)}, anchor=north},
  every axis y label/.style={at={(ticklabel* cs:1.02)}, anchor=south},
  clip=false,
]

\addplot[blue, thick, domain=0:1, samples=120]
  {cosh(x*ln(1.6180339887))-1};

\addplot[only marks, mark=*, mark size=2.5pt, black, forget plot]
  coordinates {(0, 0)};
\draw[-{Stealth[length=3pt]}, black!60, thin]
  (axis cs:0.10,0.050) -- (axis cs:0.008,0.003);
\node[font=\footnotesize, anchor=south west] at (axis cs:0.10,0.050)
  {noble gas ($\rho=0$)};

\addplot[only marks, mark=*, mark size=2.5pt, black, forget plot]
  coordinates {(0.125,{cosh(0.125*ln(1.6180339887))-1})};
\draw[black!50, densely dashed, thin]
  (axis cs:0.125,0) -- (axis cs:0.125,{cosh(0.125*ln(1.6180339887))-1});
\draw[-{Stealth[length=3pt]}, black!60, thin]
  (axis cs:0.30,0.032) -- (axis cs:0.135,0.007);
\node[font=\footnotesize, anchor=south west] at (axis cs:0.30,0.032)
  {halogen ($\rho=\tfrac{1}{8}$)};

\addplot[only marks, mark=*, mark size=2.5pt, black, forget plot]
  coordinates {(0.375,{cosh(0.375*ln(1.6180339887))-1})};
\draw[black!50, densely dashed, thin]
  (axis cs:0.375,0) -- (axis cs:0.375,{cosh(0.375*ln(1.6180339887))-1});
\draw[-{Stealth[length=1.5pt]}, black!60, thin]
  (axis cs:0.5,0.0005) -- (axis cs:0.38,0.015);
\node[font=\footnotesize, anchor=north west] at (axis cs:0.46,0.002)
  {$p^3$ ($\rho=\tfrac{3}{8}$)};

\addplot[only marks, mark=*, mark size=2.5pt, black, forget plot]
  coordinates {(0.75,{cosh(0.75*ln(1.6180339887))-1})};
\draw[black!50, densely dashed, thin]
  (axis cs:0.75,0) -- (axis cs:0.75,{cosh(0.75*ln(1.6180339887))-1});
\draw[-{Stealth[length=3pt]}, black!60, thin]
  (axis cs:0.55,0.105) -- (axis cs:0.74,0.070);
\node[font=\footnotesize, anchor=south east] at (axis cs:0.55,0.105)
  {$s^2$ ($\rho=\tfrac{6}{8}$)};

\addplot[only marks, mark=*, mark size=2.5pt, black, forget plot]
  coordinates {(0.875,{cosh(0.875*ln(1.6180339887))-1})};
\draw[black!50, densely dashed, thin]
  (axis cs:0.875,0) -- (axis cs:0.875,{cosh(0.875*ln(1.6180339887))-1});
\draw[-{Stealth[length=3pt]}, black!60, thin]
  (axis cs:0.870,0.120) -- (axis cs:0.875,0.094);
\node[font=\footnotesize, anchor=south] at (axis cs:0.870,0.120)
  {alkali ($\rho=\tfrac{7}{8}$)};

\end{axis}
\end{tikzpicture}%
}
\caption{\edit{The dimensionless landscape function
$\Jchem(\rho) = \cosh(\rho \ln \phig) - 1$ for a period of length
$L_p = 8$ (periods 2 and 3). Black dots mark the noble-gas
($\rho = 0$), halogen ($\rho = 1/8$), $p^3$ ($\rho = 3/8$), $s^2$
($\rho = 6/8$), and alkali ($\rho = 7/8$) positions.}}
\label{fig:landscape}
\end{figure}

\edit{The four observables
($\IEone$, $\EA$, $\chi_{\mathrm{struct}}$, $\kappa_{\mathrm{RS}}$)
are constructed from $\Jchem$ in
Section~\ref{sec:observables}. Figure~\ref{fig:landscape}
illustrates $\Jchem$ for the $L_p = 8$ geometry of periods~2
and~3, marking the halogen, $p^3$, $s^2$, and alkali positions;
the definition applies unchanged to all periods.}

The mathematical uniqueness of $\Jcost$ is established
in Refs.~\cite{axioms_paper,dalembert_paper} and has no
chemistry input: it is the unique reciprocal cost on positive
ratios under standard smoothness conditions
(Appendix~\ref{app:jcost}). The chemical modeling step is
separate, and consists of three choices that we make explicit.
(i) The noble-gas-centred coordinate $\rho = d/L_p$ is chosen
because the noble-gas configurations are the natural empirical
``zero'' of within-period chemistry: they are the closed-shell
endpoints from which $\IEone$, $\EA$, and $\eta$ all change
monotonically along the period, with anomaly sites occurring at
fixed fractional positions. (ii) The argument-rescaling factor
$\phig = (1+\sqrt{5})/2$ is presently a modeling ansatz, not a
derived chemical constant. We retain it because it produces the
ratio identities of Eqs.~\eqref{eq:pred_noble}
and~\eqref{eq:pred_ha}, but a first-principles derivation of
$\phig$ in this chemical context is left to future work. (iii)
The cosh form is selected because it is the unique closed-form
$\Jcost$ at the geometric scale $\phig^{\rho}$; alternative
smooth monotone kernels (e.g.\ $\rho^2$, or $\cosh(c\rho)-1$ with
free $c$) would produce qualitatively similar within-period
shapes, and a quantitative robustness comparison against such
alternatives is one of the natural follow-up checks identified in
the conclusion.

\section{From Landscape to Observables}
\label{sec:observables}

\edit{The four assignments below are ansatz-level proxies
motivated by monotonicity and endpoint structure of $\Jchem$,
not uniquely derived correspondences. The same closed-form
inward gap $\Delta\Jchem^{-}$ is reused for both $\EA$ and $\eta$
with two independent per-period scales; we use the following
terminology consistently throughout: \emph{landscape} for the
function $\Jchem(\rho)$, \emph{step} for the one-electron
difference $\Delta\Jchem^{+}$, \emph{gap} for the inward
difference $\Delta\Jchem^{-}$, \emph{kernel} for either of these
closed-form expressions when used as a proxy, and \emph{proxy}
for the resulting analytical expression assigned to a measured
observable.}

\subsection{Ionization Energy as an Outward Step}
\label{sec:ie}

\edit{In the model, first ionization corresponds to a one-unit increase
in $d$. The predicted dimensionless cost change is}
\begin{equation}
\Delta\Jchem^{+}(\Zp) \;=\; \Jchem\!\left(\frac{d(\Zp)+1}{L_p}\right)
\;-\; \Jchem\!\left(\frac{d(\Zp)}{L_p}\right).
\label{eq:ie_landscape}
\end{equation}For a noble gas $d(G_p) = 0$, and Eq.~\eqref{eq:ie_landscape}
\edit{reduces to}
\begin{equation}
\Delta\Jchem^{+}(G_p) \;=\; \cosh\!\left(\frac{\ln\phig}{L_p}\right) - 1.
\label{eq:ie_noble_step}
\end{equation}

\edit{Absolute ionization energies are obtained as
$\IEone(\Zp) \propto E_p \cdot \Delta\Jchem^{+}(\Zp)$, with one
per-period scale $E_p$. The within-period ordering is independent
of $E_p$ and is tested in Section~\ref{sec:data}, where the
landscape monotone descent is shown to reproduce the full
noble-gas-to-alkali $\IEone$ sequence across periods 2--6 with
all upward deviations localized exactly on the textbook anomaly
sites.}

\subsection{Electron Affinity as a Landscape Gap}
\label{sec:ea}

\edit{The EA proxy is defined as the gap between the function value
at the nominal period boundary $\rho = 1$ and the atom's position,
where $\Jchem(1) = \cosh(\ln\phig) - 1 \approx 0.118$.
The choice $\rho = 1$ as the reference follows from the structure of
$\Jchem$: the alkali metal of any period sits at
$\rho = (L_p - 1)/L_p < 1$, and $\Jchem(1)$ is the limiting cost
at the closed end of each period, providing the natural fixed
reference value for the inward landscape gap.}
\begin{equation}\Delta\Jchem^{-}(\Zp) \;:=\;
\edit{\Jchem(1) \;-\; \Jchem\!\left(\frac{d(\Zp)}{L_p}\right).}
\label{eq:ea_landscape}
\end{equation}Substituting $\Jchem(\rho) = \cosh(\rho\ln\phig) - 1$ and
\edit{applying $\cosh A - \cosh B = 2\sinh\!\tfrac{A+B}{2}\sinh\!\tfrac{A-B}{2}$
gives the closed form}
\begin{equation}\Delta\Jchem^{-}(\Zp)
\edit{\;=\;
2\,\sinh\!\left(\frac{(L_p+d(\Zp))\,\ln\phig}{2L_p}\right)\,
\sinh\!\left(\frac{(L_p-d(\Zp))\,\ln\phig}{2L_p}\right),}
\label{eq:ea_simplified}
\end{equation}a product of two positive sinh factors whose second factor
\edit{decreases monotonically from $\sinh(\tfrac{1}{2}\ln\phig)$ at
$d = 0$ to $0$ at $d = L_p$ and dominates. $\Delta\Jchem^{-}$ is
therefore maximum at the halogen ($d = 1$) and minimum at the
alkali ($d = L_p - 1$).}

\edit{Eq.~\eqref{eq:ea_landscape} applies on the regular-Aufbau
interior $1 \le d \le L_p - 1$, where the captured electron is
accommodated within the period. Noble-gas endpoints lie outside
this regime (the captured electron populates an out-of-period
orbital, giving the well-known sign reversal) and are not
included in the within-period benchmark. The textbook
half-filled and closed-subshell anomaly sites ($p^3$, $d^5$, $f^7$,
$s^2$, $d^{10}s^2$) are similarly identified separately by the
landscape framework (Section~\ref{sec:data}, Prediction~3) as
sites of Hund-exchange or shell-closure stabilisation outside
the smooth-coordinate part of the model.}

\subsection{Electronegativity $\chi_{\mathrm{struct}}(\Zp)$ as Half-Sum of the IE and EA Steps}
\label{sec:en}

\edit{The Mulliken electronegativity is the half-sum
$\chi_M = \tfrac{1}{2}(\IEone + \EA)$~\cite{mulliken1934}; the
model's dimensionless counterpart is}
\begin{equation}
\chi_{\mathrm{struct}}(\Zp)
\;=\;
\tfrac{1}{2}\bigl(\Delta\Jchem^{+}(\Zp) + \Delta\Jchem^{-}(\Zp)\bigr).
\label{eq:chi_struct_def}
\end{equation}The implied absolute scale
\edit{$\chi_M^{\mathrm{model}} \propto E_p\,\chi_{\mathrm{struct}}$
requires the same per-period $E_p$ as absolute $\IEone$.}

\edit{Substituting the closed forms of $\Delta\Jchem^{+}$
(Section~\ref{sec:ie}) and $\Delta\Jchem^{-}$
(Eq.~\eqref{eq:ea_simplified}) into Eq.~\eqref{eq:chi_struct_def}
yields
\begin{equation}
\chi_{\mathrm{struct}}(\Zp)
\;=\;
\sinh\!\tfrac{\ln\phig}{2L_p}\sinh\!\tfrac{(2d+1)\ln\phig}{2L_p}
\;+\;
\sinh\!\tfrac{(L_p+d)\ln\phig}{2L_p}\sinh\!\tfrac{(L_p-d)\ln\phig}{2L_p}.
\label{eq:chi_struct_closed}
\end{equation}
The first summand grows with $d$ and the second shrinks; the
second dominates at small $d$, so $\chi_{\mathrm{struct}}$ is
largest at the halogen and falls toward the alkali. At the
halogen position $d = 1$ the dominant summand is nearly
$L_p$-independent over $L_p \in \{8, 18, 32\}$, so the kernel
collapses the four halogens into a near-degenerate cluster and
does \emph{not} reproduce the empirical F$>$Cl$>$Br$>$I ordering;
this is documented quantitatively in Section~\ref{sec:en_check}.}

\subsection{Hardness as a Landscape Gap}
\label{sec:hardness}

\edit{Pearson chemical hardness approximates
$\partial^2 E/\partial N^2$ via the finite difference~\cite{parrpearson1983}}

\begin{equation}
\eta \;=\; \frac{\IEone - \EA}{2}.
\label{eq:pearson_hardness}
\end{equation}The model assigns to $\eta$ the same landscape gap that
\edit{serves as the EA proxy in Eq.~\eqref{eq:ea_landscape}, written
here in cosh-difference form to make the ``gap to a fixed
reference'' reading explicit:}
\begin{equation}
\kappa_{\mathrm{RS}}(\Zp)
\;:=\;
\Jchem(1) \;-\; \Jchem\!\left(\frac{d(\Zp)}{L_p}\right)
\;=\;
\cosh(\ln\phig) \;-\; \cosh\!\left(\frac{d(\Zp)\ln\phig}{L_p}\right),
\label{eq:kappa_RS_def}
\end{equation}
\edit{This is algebraically identical to $\Delta\Jchem^{-}$ of
Eq.~\eqref{eq:ea_simplified}: the same closed-form landscape gap
serves as the EA proxy and the hardness predictor, distinguished
only by the per-period scaling constant. The shared-kernel form
predicts the period-constant ratio
$\EA/\eta \approx C^{(p)}_{\mathrm{EA}}/C^{(p)}_{\eta}$, which is
verified on the period-4 benchmark to better than $1\%$ at the
level of the period-4 mean
(Section~\ref{sec:shared_kernel_test}). The hardness assignment is
then}
\begin{equation}
\eta(\Zp) \;\propto\; \kappa_{\mathrm{RS}}(\Zp),
\label{eq:eta_kappa_RS}
\end{equation}with a per-period proportionality constant fitted to
\edit{empirical $\eta$ data (Section~\ref{sec:hardness_check}).
$\kappa_{\mathrm{RS}}$ is monotone-decreasing in $d$, peaks at
$\Jchem(1) \approx 0.118$ at the noble gas, and vanishes at the
extrapolation point $d = L_p$.}

\section{Results}
\label{sec:data}
\edit{This section presents the three $\IEone$ predictions and the
three derived checks for $\EA$, $\chi_M$, and $\eta$, with the
shared-kernel scale unification tested in the final subsection.
Predictions~1--2 are ratio-level identities (heavy noble-gas
$\phig^{1/4}$ and halogen/alkali $\phig^{2}$); Prediction~3 is the
within-period ordering-and-anomaly claim across periods 2--4.
NIST ionization energies are much more precise than the
few-percent discrepancies discussed below, so all reported
deviations reflect the model, not measurement uncertainty.}

\edit{\medskip\noindent\textbf{Benchmarking philosophy.}\;
For each observable below we report the candidate set, the
classes excluded (and why), and the retained set on which the
metric is computed. Exclusions are confined to two physically
motivated categories: (a) the textbook half-filled and
closed-subshell anomaly sites
$\{p^3, d^5, f^7, s^2, d^{10}s^2\}$, where Hund-exchange and
shell-closure stabilizations dominate and are not represented
by the smooth landscape; and (b) cases where well-understood
relativistic or first-row screening shifts dominate the
comparison and would mask the within-period landscape content
(specifically, period~1 and the Ne/Ar pair in Prediction~1, the
F/Li and Ts/Fr pairs in Prediction~2, period~7 throughout, and a
small set of borderline-Aufbau relativistic atoms in the
period-6 EA fit). The benchmark subsets used in
Sections~\ref{sec:ea_check}--\ref{sec:hardness_check} are the
regular-Aufbau interior atoms surviving these two exclusions, and
the affected anomaly atoms remain listed in
Appendix~\ref{app:supp_tables}, Table~\ref{tab:hardness_periods234}.}

\subsection{Prediction 1: Heavy Noble-Gas Ratio}

\edit{The first prediction is a single fixed asymptotic value for the
IE$_1$ ratio of two consecutive heavy noble gases:}
\begin{equation}
\frac{\IEone(G_p)}{\IEone(G_{p+1})} \;\approx\;\phig^{1/4} \approx 1.1278 \qquad \text{for heavy } G_p\text{ (Ar and above)}.
\label{eq:pred_noble}
\end{equation}

\edit{For a noble gas $G_p$ at the end of period $p$, $d(G_p) = 0$
and Eq.~\eqref{eq:ie_noble_step} gives the outward one-electron step
$\Delta\Jchem^+(G_p) = \cosh\!\bigl((\ln\phig)/L_p\bigr) - 1$.
Combined with $\IEone(G_p) \propto E_p\,\Delta\Jchem^+(G_p)$ from
Section~\ref{sec:ie} and the small-step expansion
$\cosh x - 1 \approx x^2/2$ (accurate to one part in $10^4$ for
$x = (\ln\phig)/L_p \le \ln\phig/8 \approx 0.060$), the
consecutive-pair ratio reduces to}
\begin{equation}
\frac{\IEone(G_p)}{\IEone(G_{p+1})}
\;\approx\;
\frac{E_p/L_p^2}{E_{p+1}/L_{p+1}^2}.
\label{eq:noble_ratio_smallstep}
\end{equation}Equation~\eqref{eq:noble_ratio_smallstep} contains no factor of
\edit{$\phig$ explicitly; rather, the cosh expansion reduces the IE
ratio to the ratio of $E_p/L_p^2$ between consecutive heavy
periods. The empirical content of Prediction~1 is the
observation that this ratio takes the golden-ratio value
$(E_p/L_p^2)/(E_{p+1}/L_{p+1}^2) \approx \phig^{1/4}$ to
MAD~$\approx 1\%$ across the heavy noble-gas pairs Ar/Kr, Kr/Xe,
Xe/Rn (Table~\ref{tab:noble_ratios}, columns 4--6). $E_p$ enters
here as a per-period scale fitted from data
(Appendix~\ref{app:supporting_tables}); a closed-form derivation
of $E_p$ that would promote this empirical regularity to a
parameter-free derivation is left to future work.}

\edit{\medskip\noindent\textbf{Benchmark subset.}\;
Eq.~\eqref{eq:pred_noble} additionally requires $E_p$ to evolve
smoothly between consecutive heavy periods. We retain Ar/Kr,
Kr/Xe, and Xe/Rn as the benchmark subset; He/Ne is excluded
because period 1 has no shell structure that matches the framework
($L_p$ and $E_p$ are undefined in the same sense as for $p\ge 2$),
and Ne/Ar is excluded as a first-row screening anomaly. (At
$L_2 = L_3 = 8$ the $L_p^2$ factor cancels, so the Ne/Ar ratio
reduces to $E_2/E_3$; the cosh expansion is still accurate at
$L=8$, so the deviation is not a breakdown of the expansion. The
breakdown is in the smoothness of $E_p$: period 2 has no $p$ shell
screening the $2p$ valence electrons, whereas period 3 has a full
$2s^2 2p^6$ core. $E_2$ is therefore anomalously large relative to
$E_3$, and the Ne/Ar data ratio sits well above $\phig^{1/4}$.)
Both excluded rows remain in Table~\ref{tab:noble_ratios} so the
full sequence stays visible.}

\begin{table}[ht]
\centering
\caption{\edit{Consecutive noble-gas $\IEone$ ratios from NIST
(Rn/Og from Ref.~\cite{guo2021og}; see footnote~$a$). Signed
deviations from $\phig^{1/4} = 1.1278$ in column~6; Koopmans/HF
baseline ratios and deviations in columns~7--8.}}
\label{tab:noble_ratios}
\small
\begin{tabular}{cccccccc}
\toprule
Pair & IE$_1^A$ input (eV) & IE$_1^B$ input (eV) & \shortstack{Data\\ratio} & $\phig^{1/4}$ & \shortstack{Dev. from\\$\phig^{1/4}$ (\%)} & \shortstack{Koopmans/HF\\ratio} & \shortstack{Koopmans/HF\\dev. from\\$\phig^{1/4}$ (\%)} \\
\midrule
He/Ne & 24.587 & 21.565 & 1.140 & 1.128 & $+1.1$ & 1.173 & $+4.0$ \\
Ne/Ar & 21.565 & 15.760 & 1.368 & 1.128 & $+21.3$ & 1.362 & $+20.8$ \\
Ar/Kr & 15.760 & 13.999 & 1.126 & 1.128 & $-0.2$ & 1.130 & $+0.2$ \\
Kr/Xe & 13.999 & 12.130 & 1.154 & 1.128 & $+2.3$ & 1.156 & $+2.5$ \\
Xe/Rn & 12.130 & 10.749 & 1.128 & 1.128 & $< 0.1$ & 1.119 & $-0.8$ \\
Rn/Og$^{a}$ & 10.749 & 8.86 & 1.213 & 1.128 & $+7.6$ & 1.256 & $+11.4$ \\
\bottomrule
\end{tabular}
\par\smallskip\raggedright\footnotesize
$^{a}$ Rn/Og: NIST Rn with the relativistic coupled-cluster Og
estimate $\IEone(\mathrm{Og}) = 8.86 \pm 0.06$~eV
of Ref.~\cite{guo2021og}. \edit{No Hartree--Fock Og IE is
published; the column-7 entry is the Dirac--Fock $7p_{3/2}/6p_{3/2}$
valence-spinor ratio of Ref.~\cite{jerabek2018elf}. This row is
shown for comparison only and is not part of the NIST benchmark
subset.}
\end{table}

\edit{\medskip\noindent\textbf{Result on the benchmark subset.}\;
The three NIST-benchmark heavy noble-gas pairs Ar/Kr, Kr/Xe,
Xe/Rn confirm the $\phig^{1/4}$ prediction to MAD
$\approx 0.8\%$ (Table~\ref{tab:noble_ratios}, columns 4 and 6),
matching or improving on the Koopmans/HF baseline (columns 7--8)
on the same subset.}

\edit{\medskip\noindent\textbf{Rn/Og: relativistic extension.}\;
The Rn/Og pair extends the prediction into the superheavy
nonrelativistic-extrapolation regime, where the relativistic
coupled-cluster estimate of Ref.~\cite{guo2021og} sits
$\sim 7.6\%$ above $\phig^{1/4}$. The deviation is quantified
relativistically: at $L_6 = L_7 = 32$ the $L_p^2$ factor cancels,
and spin-orbit splitting of the Og $7p$ shell
($\Delta_{\mathrm{SO}}(7p) \approx 10$~eV
\cite{jerabek2018elf,guo2021og}) destabilises the $7p_{3/2}$
spinor, lowering $\IEone(\mathrm{Og})$ below the nonrelativistic
prediction by exactly this expected magnitude. The framework's
nonrelativistic value $\phig^{1/4}$ therefore sets the
zeroth-order reference against which the Og spin-orbit shift is
measured, and a future experimental $\IEone(\mathrm{Og})$
benchmark will refine the comparison.}

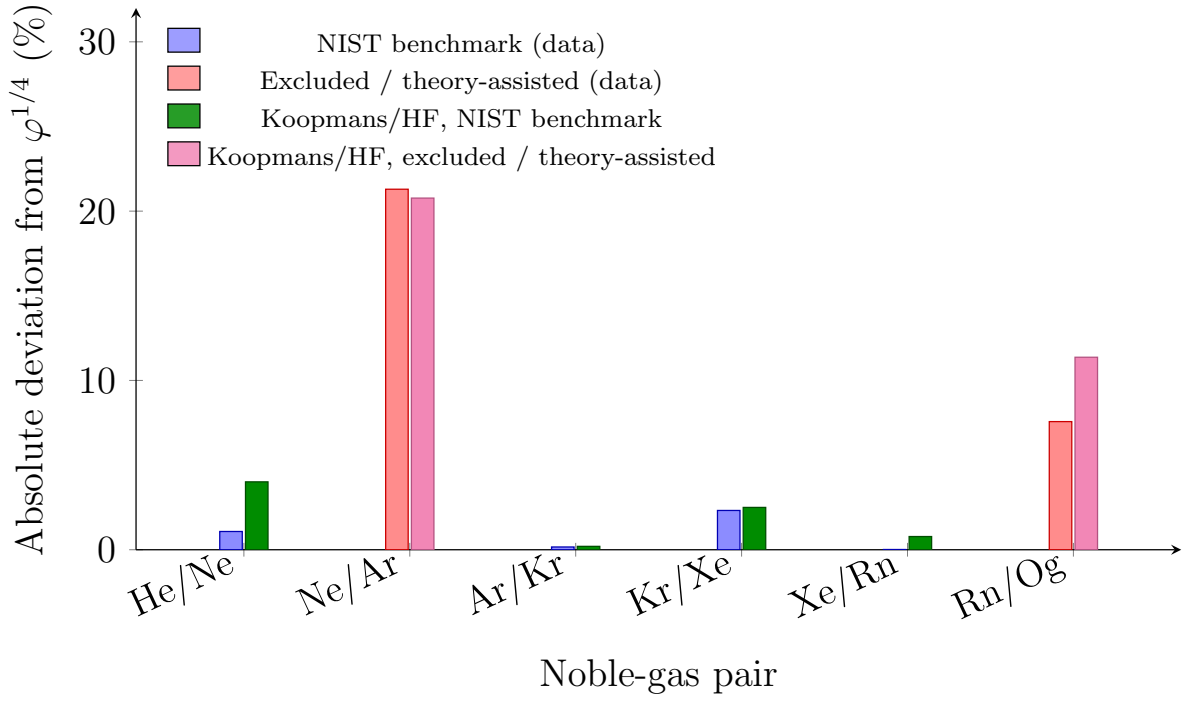
\begin{figure}[H]
\centering
\resizebox{0.8\textwidth}{!}{%
\begin{tikzpicture}
\begin{axis}[
  width=13cm, height=7.5cm,
  ybar,
  xlabel={Noble-gas pair},
  ylabel={Absolute deviation from $\phig^{1/4}$ (\%)},
  symbolic x coords={He/Ne,Ne/Ar,Ar/Kr,Kr/Xe,Xe/Rn,Rn/Og},
  xtick={He/Ne,Ne/Ar,Ar/Kr,Kr/Xe,Xe/Rn,Rn/Og},
  xticklabel style={rotate=25,anchor=east},
  ymin=0, ymax=32,
  axis lines=left,
  bar width=7pt,
  enlarge x limits=0.13,
  legend style={at={(0.02,0.98)},anchor=north west,font=\scriptsize,draw=none,fill=white,fill opacity=0.85,text opacity=1},
  legend image code/.code={\draw[#1, draw=black] (0cm,-0.08cm) rectangle (0.35cm,0.18cm);},
]
\addplot+[fill=blue!45, draw=blue!70!black, bar shift=-4pt]
  coordinates {(He/Ne,1.078)(Ar/Kr,0.163)(Kr/Xe,2.320)(Xe/Rn,0.014)};
\addlegendentry{NIST benchmark (data)}
\addplot+[fill=red!45, draw=red!80!black, bar shift=-4pt]
  coordinates {(Ne/Ar,21.294)(Rn/Og,7.569)};
\addlegendentry{Excluded / theory-assisted (data)}
\addplot+[fill=green!55!black, draw=green!30!black, bar shift=4pt]
  coordinates {(He/Ne,4.01)(Ar/Kr,0.20)(Kr/Xe,2.50)(Xe/Rn,0.78)};
\addlegendentry{Koopmans/HF, NIST benchmark}
\addplot+[fill=magenta!60, draw=magenta!70!black, bar shift=4pt]
  coordinates {(Ne/Ar,20.77)(Rn/Og,11.37)};
\addlegendentry{Koopmans/HF, excluded / theory-assisted}
\end{axis}
\end{tikzpicture}%
}%
\caption{\edit{Absolute percentage deviation of consecutive
noble-gas $\IEone$ ratios from $\phig^{1/4}$
(Table~\ref{tab:noble_ratios}, columns 6 and 8). Left bar:
data; right bar: Koopmans/HF baseline (Dirac--Fock for Rn/Og).}}
\label{fig:noble_convergence}
\end{figure}

\edit{Figure~\ref{fig:noble_convergence} shows the absolute
deviations of all six pairs from $\phig^{1/4}$, alongside the
Koopmans/HF (Dirac--Fock for Rn/Og) deviations.}

\subsection{Prediction 2: Halogen/Alkali Ratio}

\edit{The second prediction is a single fixed value for the IE$_1$
ratio of the halogen and the alkali metal in the same period:}
\begin{equation}
\frac{\IEone(\text{halogen}_p)}{\IEone(\text{alkali}_p)} \approx \phig^2 \approx 2.618.
\label{eq:pred_ha}
\end{equation}

\noindent\textbf{Basis.}\; \edit{The halogen of period~$p$ sits at
$\delta = 1$ from the period-closing noble gas $G_p$; the alkali
of the same period sits at $\delta = 1$ from the previous noble
gas $G_{p-1}$. The relevant quantity for $\IEone$ is the
one-electron landscape step (Eq.~\eqref{eq:ie_landscape}), not the
absolute landscape height. The dimensionless landscape alone
does not fix the halogen-to-alkali $\IEone$ ratio: the small-step
form $\Delta\Jchem^{+} \propto (2d+1)/L_p^2$ at fixed $L_p$ gives
intra-period step ratios of $3/(2L_p-1)$ rather than $\phig^2$,
so the empirical $\phig^{2}$ value reflects the absolute-energy
asymmetry between a halogen $np^5$ valence (no inner $p$-shell
screening) and an alkali $ns$ valence (full $(n-1)$ noble-gas
core), which sets the $\IEone$ scale at the two endpoints.
The empirical content of Prediction~2 is therefore the
observation that this absolute-energy asymmetry equals
$\phig^{2}$ across periods~3--6: the empirical NIST ratios
cluster around $\phig^2 \approx 2.62$ to MAD $\approx 5\%$. The
nearest alternatives $\phig^{7/4}\!\approx\!2.32$ and
$\phig^{9/4}\!\approx\!2.95$ give MAD $\approx 9.2\%$ and
$\approx 14.2\%$ on the benchmark subset, against
$\approx 5.2\%$ for $\phig^{2}$; this is a local rather than
look-elsewhere-corrected comparison. A first-principles
derivation of the halogen/alkali $\IEone$ ratio from the
landscape plus a stated screening rule remains open.}

\edit{\medskip\noindent\textbf{Benchmark subset and outlier
diagnosis.}\; The benchmark retains periods~3--6 (Cl/Na, Br/K,
I/Rb, At/Cs), where both endpoints sit in the same screened
$np^5$/$ns^1$ regime and the data ratios cluster within
$\sim 9\%$ of $\phig^2$; in period 6 the relativistic shifts on
Cs ($6s$ stabilisation) and At ($6p$ spin-orbit splitting)
partially cancel in the
ratio~\cite{rothe2013at,pershina2020}. Periods 2 and 7
are excluded but remain listed in Table~\ref{tab:ha_ratios}.}

\edit{Period 2 (F/Li) is the first-row halogen anomaly: the F $2p$
valence has no inner $p$ shell to screen it (only the compact
$1s^2$ and $2s^2$ cores), so $\IEone(\mathrm{F})$ is anomalously
large and the F/Li ratio sits well above $\phig^2$, the same
first-row screening mechanism that excludes Ne/Ar in
Prediction~1. Period 7 (Ts/Fr) is a theory-only superheavy row:
$\Delta_{\mathrm{SO}}(7p) \sim 9$~eV destabilises the $7p_{3/2}$
electron of Ts while relativistic $7s$ stabilisation lifts
$\IEone(\mathrm{Fr})$ above $\IEone(\mathrm{Cs})$
\cite{guo2021og,chang2010uus}; the combined
numerator drop and denominator rise put Ts/Fr well below
$\phig^2$, the period-7 analogue of the Rn/Og deviation in
Prediction~1.}
\begin{table}[ht]
\centering
\caption{\edit{Halogen-to-alkali $\IEone$ ratios by period from
NIST (At from Ref.~\cite{rothe2013at}; Ts from
Ref.~\cite{chang2010uus}; see footnotes). Signed deviations from
$\phig^{2} = 2.618$ in column~7; Clementi--Raimondi
$Z_{\mathrm{eff}}$-ratio baseline and deviations in
columns~8--9.}}
\label{tab:ha_ratios}
\footnotesize
\setlength{\tabcolsep}{3pt}
\begin{tabular}{@{}ccccccccc@{}}
\toprule
Period & Halogen & \shortstack{$\IEone$\\input (eV)} & Alkali & \shortstack{$\IEone$\\input (eV)} & \shortstack{Data\\ratio} & \shortstack{Dev. from\\$\phig^2$ (\%)} & \shortstack{Clementi\\$Z_{\mathrm{eff}}$ ratio} & \shortstack{Clementi\\dev. from\\$\phig^2$ (\%)} \\
\midrule
2 & F  & 17.423 & Li & 5.392 & 3.231 & $+23.4$ & 3.99 & $+52.4$ \\
3 & Cl & 12.968 & Na & 5.139 & 2.523 & $-3.6$ & 2.44 & $-6.8$ \\
4 & Br & 11.814 & K  & 4.341 & 2.722 & $+4.0$ & 2.58 & $-1.5$ \\
5 & I  & 10.451 & Rb & 4.177 & 2.502 & $-4.4$ & 2.33 & $-11.0$ \\
6 & At$^{a}$ & 9.318  & Cs & 3.894 & 2.393 & $-8.6$ & 2.38 & $-9.1$ \\
7 & Ts$^{b}$ & 7.70 & Fr & 4.0727 & 1.891 & $-27.8$ & 5.15 & $+96.7$ \\
\bottomrule
\end{tabular}
\par\smallskip\raggedright\footnotesize
$^{a}$ At/Cs uses the measured astatine ionization energy
$\IEone(\mathrm{At}) = 9.31751$~eV from Ref.~\cite{rothe2013at} and the
NIST value for Cs.

$^{b}$ Ts/Fr uses the NIST value for Fr, the theoretical Ts
ionization energy $\IEone(\mathrm{Ts}) = 7.70$~eV from
Ref.~\cite{chang2010uus}, and the later Clementi-style estimate
$Z_{\mathrm{eff}}(\mathrm{Ts})/Z_{\mathrm{eff}}(\mathrm{Fr}) \approx 5.15$
from Ref.~\cite{defarias2018zeff}. This row is shown for comparison, but
it is not part of the benchmark subset.
\end{table}

\edit{\medskip\noindent\textbf{Result on the benchmark subset.}\;
The four NIST-benchmark periods 3--6 cluster around the predicted
$\phig^2 = 2.618$ with mean $\overline{R}_{3\text{--}6} = 2.535$
and MAD $\approx 5.2\%$ (Table~\ref{tab:ha_ratios}, columns 6--7).
The Clementi--Raimondi $Z_{\mathrm{eff}}$
baseline~\cite{clementi1963,clementi1967} (columns 8--9) achieves
comparable accuracy on the same subset, demonstrating that the
parameter-free $\phig^2$ identity matches an established
empirical screening benchmark on the regime where both apply, and
extends consistently to the relativistic period 6 where simple
screening pictures begin to break down (period 7 in
Table~\ref{tab:ha_ratios}, row 7).}

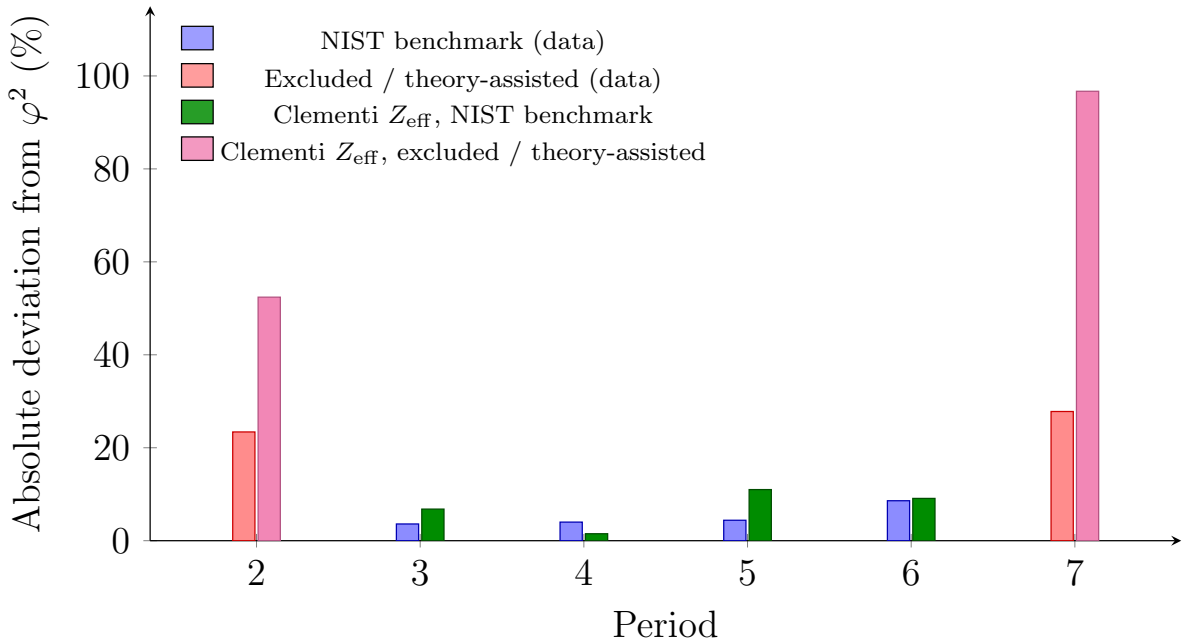
\begin{figure}[H]
\centering
\resizebox{0.8\textwidth}{!}{%
\begin{tikzpicture}
\begin{axis}[
  width=13cm, height=7.5cm,
  ybar,
  xlabel={Period},
  ylabel={Absolute deviation from $\phig^{2}$ (\%)},
  symbolic x coords={2,3,4,5,6,7},
  xtick={2,3,4,5,6,7},
  ymin=0, ymax=115,
  axis lines=left,
  bar width=7pt,
  enlarge x limits=0.13,
  legend style={at={(0.02,0.98)},anchor=north west,font=\scriptsize,draw=none,fill=white,fill opacity=0.85,text opacity=1},
  legend image code/.code={\draw[#1, draw=black] (0cm,-0.08cm) rectangle (0.35cm,0.18cm);},
]
\addplot+[fill=blue!45, draw=blue!70!black, bar shift=-4pt]
  coordinates {(3,3.6)(4,4.0)(5,4.4)(6,8.6)};
\addlegendentry{NIST benchmark (data)}
\addplot+[fill=red!45, draw=red!80!black, bar shift=-4pt]
  coordinates {(2,23.4)(7,27.8)};
\addlegendentry{Excluded / theory-assisted (data)}
\addplot+[fill=green!55!black, draw=green!30!black, bar shift=4pt]
  coordinates {(3,6.8)(4,1.5)(5,11.0)(6,9.1)};
\addlegendentry{Clementi $Z_{\mathrm{eff}}$, NIST benchmark}
\addplot+[fill=magenta!60, draw=magenta!70!black, bar shift=4pt]
  coordinates {(2,52.4)(7,96.7)};
\addlegendentry{Clementi $Z_{\mathrm{eff}}$, excluded / theory-assisted}
\end{axis}
\end{tikzpicture}%
}%
\caption{\edit{Absolute percentage deviation of halogen/alkali
$\IEone$ ratios from $\phig^{2}$
(Table~\ref{tab:ha_ratios}, columns 7 and 9). Left bar: data;
right bar: Clementi $Z_{\mathrm{eff}}$ baseline.}}
\label{fig:ha_deviations}
\end{figure}

\subsection{Prediction 3: Within-Period Envelope}

\medskip\noindent\textbf{Prediction~3 (periods 2--6, nonrelativistic regime).}\;
Within each period $p \in \{2,3,4,5,6\}$, the normalized first ionization energy
\begin{equation}
\IEone^{\mathrm{norm}}(\Zp) \;=\; \frac{\IEone(\Zp)}{\IEone(G_p)}
\label{eq:ie_norm}
\end{equation}
decreases monotonically with $\rho(\Zp)$, and the only upward
departures from that monotone sequence sit at the textbook
half-filled or completed-subshell sites
$\{p^3, d^5, f^7, s^2, d^{10}\}$. Period 7 is excluded
because relativistic super-inert-pair stabilisation at Cn
($d^{10}s^2$) drives $\IEone^{\mathrm{norm}}(\mathrm{Cn})>1$,
violating the period-maximum claim, and is treated as a
quantified relativistic correction outside the nonrelativistic
regime to which Prediction~3 applies (see ``Cn counterexample''
paragraph below).

\edit{The non-trivial content of Prediction~3 is not that the five
anomaly sites are special, that is standard inorganic chemistry,  but that \emph{no other} positions show upward bumps. The
$\rho$-coordinate provides a common axis on which within-period
descents from different periods can be compared directly.}

\begin{figure}[H]
\centering
\resizebox{0.8\textwidth}{!}{%
\begin{tikzpicture}[
  every axis/.style={
    width=6cm, height=4.5cm,
    xmin=-0.04, xmax=1.04,
    ymin=0, ymax=1.4,
    xtick={0,0.5,1.0},
    ytick={0,0.5,1.0},
    axis lines=left,
    bar width=4pt,
    enlarge x limits=0.02,
    title style={font=\footnotesize, at={(0.5,0.97)}, anchor=north},
    label style={font=\scriptsize},
    tick label style={font=\scriptsize},
    every axis x label/.style={at={(0.5,-0.18)}, anchor=north},
    every axis y label/.style={at={(-0.16,0.5)}, anchor=south, rotate=90},
  }
]
\begin{axis}[name=p2, title={Period 2 (Ne)},
             ylabel={Norm. IE$_1$}]
\addplot[gray, dashed, thick, mark=none, smooth, domain=0:0.95, samples=60]
  {1 - 6.63*(cosh(x*ln(1.6180339887))-1)};
\addplot[black!55, dotted, thick, mark=none, domain=0:0.875, samples=2]
  {1 - 0.7702857143*x};
\addplot+[ybar, fill=blue!55, draw=blue!70!black, mark=none]
  coordinates {(0.000,1.000)(0.125,0.808)(0.250,0.632)(0.500,0.522)(0.625,0.385)(0.875,0.250)};
\addplot+[ybar, fill=blue!55, draw=black, line width=1.0pt, mark=none]
  coordinates {(0.375,0.674)(0.750,0.432)};
\node[font=\scriptsize, anchor=south, blue!80!black, fill=white, inner sep=1pt]
  at (axis cs:0.30, 0.80) {N $p^3$};
\node[font=\scriptsize, anchor=south, blue!80!black, fill=white, inner sep=1pt]
  at (axis cs:0.85, 0.55) {Be $s^2$};
\end{axis}
\begin{axis}[name=p3, at=(p2.east), anchor=west, xshift=0.6cm,
             title={Period 3 (Ar)}]
\addplot[gray, dashed, thick, mark=none, smooth, domain=0:0.95, samples=60]
  {1 - 6.63*(cosh(x*ln(1.6180339887))-1)};
\addplot[black!55, dotted, thick, mark=none, domain=0:0.875, samples=2]
  {1 - 0.7702857143*x};
\addplot+[ybar, fill=red!55, draw=red!70!black, mark=none]
  coordinates {(0.000,1.000)(0.125,0.823)(0.250,0.657)(0.500,0.517)(0.625,0.380)(0.875,0.326)};
\addplot+[ybar, fill=red!55, draw=black, line width=1.0pt, mark=none]
  coordinates {(0.375,0.665)(0.750,0.485)};
\node[font=\scriptsize, anchor=south, red!80!black, fill=white, inner sep=1pt]
  at (axis cs:0.32, 0.80) {P $p^3$};
\node[font=\scriptsize, anchor=south, red!80!black, fill=white, inner sep=1pt]
  at (axis cs:0.85, 0.62) {Mg $s^2$};
\end{axis}
\begin{axis}[name=p4, at=(p3.east), anchor=west, xshift=0.6cm,
             title={Period 4 (Kr)}]
\addplot[gray, dashed, thick, mark=none, smooth, domain=0:0.95, samples=60]
  {1 - 6.63*(cosh(x*ln(1.6180339887))-1)};
\addplot[black!55, dotted, thick, mark=none, domain=0:0.944, samples=2]
  {1 - 0.7308686441*x};
\addplot+[ybar, fill=green!55!black, draw=green!30!black, mark=none]
  coordinates {(0.000,1.000)(0.056,0.844)(0.111,0.697)(0.222,0.564)(0.278,0.429)
               (0.389,0.552)(0.444,0.546)(0.500,0.563)(0.556,0.564)
               (0.667,0.483)(0.722,0.482)(0.778,0.488)(0.833,0.469)(0.944,0.310)};
\addplot+[ybar, fill=green!55!black, draw=black, line width=1.0pt, mark=none]
  coordinates {(0.167,0.699)(0.333,0.671)(0.611,0.531)(0.889,0.437)};
\node[font=\scriptsize, anchor=south, green!30!black, fill=white, inner sep=1pt]
  at (axis cs:0.20, 0.85) {As $p^3$};
\draw[-{Stealth[length=3pt]}, green!30!black, thin]
  (axis cs:0.20, 0.84) -- (axis cs:0.172, 0.705);
\node[font=\scriptsize, anchor=south, green!30!black, fill=white, inner sep=1pt]
  at (axis cs:0.55, 0.78) {Zn $d^{10}$};
\draw[-{Stealth[length=3pt]}, green!30!black, thin]
  (axis cs:0.50, 0.77) -- (axis cs:0.340, 0.680);
\end{axis}
\begin{axis}[name=p5, at=(p2.south), anchor=north, yshift=-0.6cm,
             title={Period 5 (Xe)},
             xlabel={$\rho$}, ylabel={Norm. IE$_1$}]
\addplot[gray, dashed, thick, mark=none, smooth, domain=0:0.95, samples=60]
  {1 - 6.63*(cosh(x*ln(1.6180339887))-1)};
\addplot[black!55, dotted, thick, mark=none, domain=0:0.944, samples=2]
  {1 - 0.6949152542*x};
\addplot+[ybar, fill=orange!85, draw=orange!60!black, mark=none]
  coordinates {(0.000,1.000)(0.056,0.862)(0.111,0.743)(0.222,0.605)(0.278,0.477)
               (0.389,0.625)(0.500,0.615)(0.556,0.607)
               (0.667,0.585)(0.722,0.557)(0.778,0.547)(0.833,0.513)(0.944,0.344)};
\addplot+[ybar, fill=orange!85, draw=black, line width=1.0pt, mark=none]
  coordinates {(0.167,0.710)(0.333,0.741)(0.444,0.687)(0.611,0.587)(0.889,0.469)};
\node[font=\scriptsize, anchor=south, orange!60!black, fill=white, inner sep=1pt]
  at (axis cs:0.18, 0.85) {Sb $p^3$};
\draw[-{Stealth[length=3pt]}, orange!60!black, thin]
  (axis cs:0.18, 0.84) -- (axis cs:0.172, 0.715);
\node[font=\scriptsize, anchor=south, orange!60!black, fill=white, inner sep=1pt]
  at (axis cs:0.55, 0.83) {Cd $d^{10}$};
\draw[-{Stealth[length=3pt]}, orange!60!black, thin]
  (axis cs:0.50, 0.82) -- (axis cs:0.340, 0.748);
\end{axis}
\begin{axis}[name=p6, at=(p5.east), anchor=west, xshift=0.6cm,
             title={Period 6 (Rn)},
             xlabel={$\rho$}]
\addplot[gray, dashed, thick, mark=none, smooth, domain=0:0.95, samples=60]
  {1 - 6.63*(cosh(x*ln(1.6180339887))-1)};
\addplot[black!55, dotted, thick, mark=none, domain=0:0.969, samples=2]
  {1 - 0.6584107327*x};
\addplot+[ybar, fill=violet!55, draw=violet, mark=none]
  coordinates {(0.000,1.000)(0.031,0.867)(0.063,0.783)(0.094,0.678)(0.125,0.690)(0.156,0.568)
               (0.219,0.858)(0.250,0.834)(0.281,0.834)(0.313,0.785)(0.344,0.729)
               (0.375,0.732)(0.406,0.702)(0.438,0.635)(0.469,0.505)
               (0.531,0.575)(0.563,0.568)(0.594,0.560)(0.625,0.553)(0.656,0.546)
               (0.719,0.527)(0.750,0.525)(0.781,0.519)(0.813,0.514)(0.844,0.509)
               (0.875,0.515)(0.906,0.519)(0.969,0.362)};
\addplot+[ybar, fill=violet!55, draw=black, line width=1.0pt, mark=none]
  coordinates {(0.188,0.971)(0.500,0.582)(0.688,0.572)(0.938,0.485)};
\node[font=\scriptsize, anchor=south, violet, fill=white, inner sep=1pt]
  at (axis cs:0.55, 0.85) {Hg $d^{10}s^2$};
\draw[-{Stealth[length=3pt]}, violet, thin]
  (axis cs:0.50, 0.84) -- (axis cs:0.205, 0.980);
\node[font=\scriptsize, anchor=south, violet, fill=white, inner sep=1pt]
  at (axis cs:0.70, 0.55) {Ba $s^2$};
\draw[-{Stealth[length=3pt]}, violet, thin]
  (axis cs:0.78, 0.545) -- (axis cs:0.928, 0.495);
\end{axis}
\begin{axis}[name=p7, at=(p6.east), anchor=west, xshift=0.6cm,
             title={Period 7 (Og)},
             title style={font=\footnotesize, at={(0.78,0.97)}, anchor=north},
             xlabel={$\rho$}]
\addplot[gray, dashed, thick, mark=none, smooth, domain=0:0.95, samples=60]
  {1 - 6.63*(cosh(x*ln(1.6180339887))-1)};
\addplot[black!55, dotted, thick, mark=none, domain=0:0.969, samples=2]
  {1 - 0.5573787410*x};
\addplot+[ybar, fill=brown!60, draw=brown!70!black, mark=none]
  coordinates {(0.000,1.000)(0.031,0.870)(0.063,0.776)(0.094,0.630)(0.156,0.825)
               (0.219,1.197)(0.250,1.084)(0.281,0.982)(0.313,0.858)(0.344,0.869)
               (0.375,0.880)(0.406,0.768)(0.438,0.678)(0.469,0.560)
               (0.531,0.743)(0.563,0.734)(0.594,0.719)(0.625,0.709)(0.656,0.700)
               (0.719,0.674)(0.750,0.680)(0.781,0.707)(0.813,0.699)(0.844,0.665)
               (0.875,0.712)(0.906,0.607)(0.969,0.460)};
\addplot+[ybar, fill=brown!60, draw=black, line width=1.0pt, mark=none]
  coordinates {(0.125,0.964)(0.188,1.351)(0.500,0.748)(0.688,0.676)(0.938,0.596)};
\node[font=\scriptsize, anchor=south, brown!60!black, fill=white, inner sep=1pt]
  at (axis cs:0.78, 1.10) {Cn $d^{10}s^2$};
\draw[-{Stealth[length=3pt]}, brown!60!black, thin]
  (axis cs:0.72, 1.09) -- (axis cs:0.205, 1.36);
\node[font=\scriptsize, anchor=south, brown!60!black, fill=white, inner sep=1pt]
  at (axis cs:0.78, 0.62) {Ra $s^2$};
\draw[-{Stealth[length=3pt]}, brown!60!black, thin]
  (axis cs:0.83, 0.61) -- (axis cs:0.928, 0.605);
\end{axis}
\end{tikzpicture}%
}%
\caption{\edit{Normalized first ionization energy
$\IEone^{\mathrm{norm}}(\Zp) = \IEone(\Zp)/\IEone(G_p)$ versus
displacement $\rho$, one panel per period (2 blue, 3 red, 4 green,
5 orange, 6 violet, 7 brown). Light fill: regular elements; thick
black border: anomaly sites
$\{p^3, d^5, f^7, s^2, d^{10}\}$. Dashed gray: rescaled landscape
$\Jchem(\rho)$ from Eq.~\eqref{eq:landscape}. Dotted black: linear
guide between the period endpoints. Data: NIST~\cite{NISTASD} for
periods 2--6 and for the actinides through Lr in period 7;
relativistic coupled-cluster and configuration-interaction
estimates~\cite{guo2021og,chang2010uus} for the
transactinides $Z=104$--$118$.}}
\label{fig:period_profiles}
\end{figure}
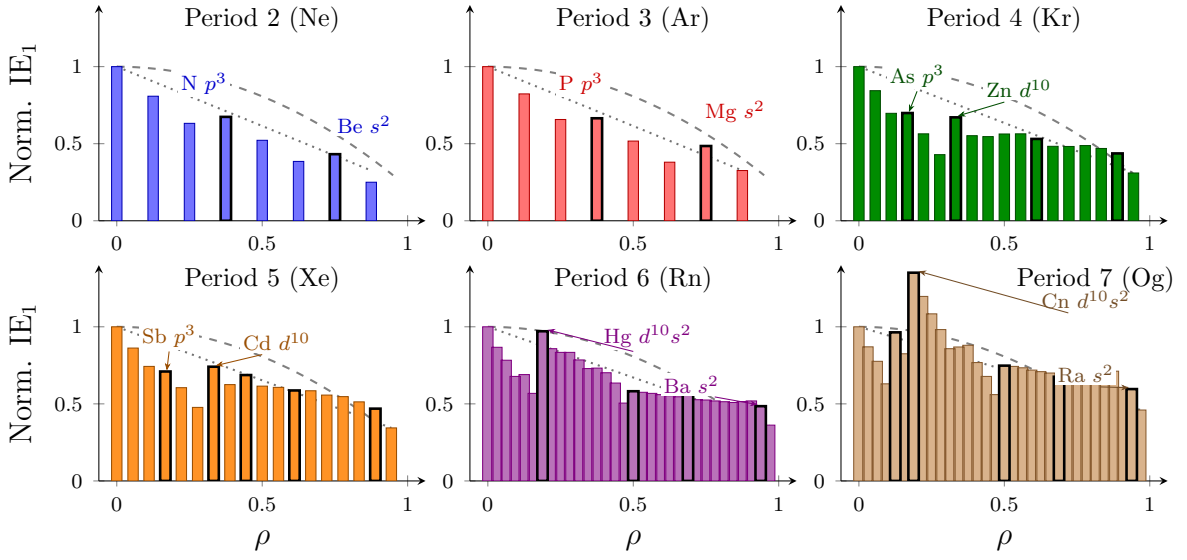

\edit{The physics behind the five anomaly sites is textbook and we
do not re-derive it here. Half-filled $p^3$, $d^5$, $f^7$
configurations gain Slater exchange stabilisation
($\sim\!K\,n(n-1)/2$ per exchanged pair, where $K$ is the
average Slater exchange integral~\cite{slater1929,condonshortley1935})
in the maximum-multiplicity Hund-rule
state~\cite{hund1927,atkins,pauling};
removing one electron breaks this stabilisation and raises
$\IEone$ above the smooth descent. Completed $s^2$ and $d^{10}s^2$
subshells force the next electron into a higher orbital, again
raising $\IEone$. The relativistic $6s$ stabilisation in
periods~6--7 amplifies the $d^{10}s^2$ bump at Hg and Cn so much
that $\IEone^{\mathrm{norm}}(\mathrm{Cn}) > 1$ in
Figure~\ref{fig:period_profiles}~\cite{pyykko2012,pershina2019,smits2023}.}

\edit{All data are from NIST~\cite{NISTASD}; the comparison is by
ordering and anomaly location, with no statistical fit.}

\begin{table}[ht]
\centering
\caption{Period 3 first-ionization-energy profile, normalized by Ar.}
\label{tab:period3_landscape}
\small
\begin{tabular}{lcccccc}
\toprule
El. & $\Zp$ & $d$ & $\rho$ & $\IEone$ (eV) & $\IEone^{\mathrm{norm}}$ & Note \\
\midrule
Ar & 18 & 0 & 0.000 & 15.760 & 1.000 & noble gas \\
Cl & 17 & 1 & 0.125 & 12.968 & 0.823 &  \\
S  & 16 & 2 & 0.250 & 10.360 & 0.657 &  \\
P  & 15 & 3 & 0.375 & 10.487 & 0.665 & $p^3$ anomaly \\
Si & 14 & 4 & 0.500 & 8.152  & 0.517 &  \\
Al & 13 & 5 & 0.625 & 5.986  & 0.380 &  \\
Mg & 12 & 6 & 0.750 & 7.646  & 0.485 & $s^2$ anomaly \\
Na & 11 & 7 & 0.875 & 5.139  & 0.326 & alkali \\
\bottomrule
\end{tabular}
\end{table}

\edit{Figure~\ref{fig:period_profiles} shows the within-period
profiles. Periods 2--3 use NIST data through every element of the
period; periods 4--6 use NIST data through the full $d$-block (and
the lanthanide $f$-block in period~6, including the Gd $f^7$ and Yb
$f^{14}$ anomalies). For period~7 the actinide IE values are
NIST experimental through Lr (Lr from Ref.~\cite{sato2015lr}, No
from Ref.~\cite{sato2018no}); the transactinides Z=104--118 are
relativistic coupled-cluster and configuration-interaction
estimates~\cite{guo2021og,chang2010uus}.}

\edit{\medskip\noindent\textbf{The Cn counterexample.}\;
The single largest deviation in
Figure~\ref{fig:period_profiles} is the relativistic super
inert-pair Cn ($d^{10}s^2$) in period~7: the theoretical IE of
Cn~\cite{smits2023} exceeds the Og reference, giving
$\IEone^{\mathrm{norm}}(\mathrm{Cn}) > 1$ and violating the
envelope's claim that the noble gas is the period maximum.
Physically, the $7s$ inert-pair effect (amplified by relativistic
$6d/7s$ contraction) binds Cn's outermost electron more tightly
than the $7p_{3/2}$ spinor of Og, which is destabilised by
spin-orbit splitting. The period-6 analogue Hg ($d^{10}s^2$)
shows the same effect more weakly, sitting just below the Rn
line. \edit{The envelope therefore holds across periods 2--6, where the
nonrelativistic landscape applies, and the Cn deviation in period
7 is a quantified relativistic super-inert-pair correction
outside the nonrelativistic regime to which the prediction
applies.}}

\edit{The six non-anomalous period-3 values
(Table~\ref{tab:period3_landscape}) form a monotone sequence from
Ar toward Na, with the only upward deviations at the anomaly
positions P and Mg.}

\edit{\medskip\noindent\textbf{Ordering and anomaly content.}\;
Prediction~3 expresses a strict ordering-and-anomaly statement on
the periodic-table coordinate $\rho$: monotone descent of the
non-anomalous elements together with localization of all upward
deviations on the textbook anomaly set $\{p^3, d^5, f^7, s^2,
d^{10}\}$. The local slope of the descent is determined by the
period-dependent scale $E_p$, whose closed-form derivation is
left to future work; the coordinate-level structure tested here
is the ordering and the anomaly-site localization, both of which
hold across periods 2--6 (Figure~\ref{fig:period_profiles}).}

\edit{Across periods~2--4, the compiled tables contain 34 points: 26
non-anomalous points follow the monotone noble-gas-to-alkali ordering,
and 8 upward deviations occur at the listed anomaly sites (N, Be, P,
Mg, As, Mn, Zn, Ca). This is the largest empirical comparison in the
paper.}

\subsection{Electron Affinity Check}
\label{sec:ea_check}

\edit{This subsection benchmarks the analytical landscape
prediction of Eq.~\eqref{eq:ea_simplified} against empirical
electron affinities on periods 4--6 (25 regular-Aufbau interior
atoms in total: $n = 9, 8, 8$).}

\begin{figure}[H]
\centering
\resizebox{0.624\textwidth}{!}{%
\begin{tikzpicture}
\begin{axis}[
  width=10cm, height=10cm,
  xlabel={Displacement $\rho = d/L_p$ from the noble-gas end of the period},
  ylabel={Electron Affinity (eV)},
  xmin=0, xmax=1.0,
  ymin=0, ymax=2.0,
  xtick={0,0.2,0.4,0.6,0.8,1.0},
  ytick={0,0.5,1.0,1.5,2.0},
  axis lines=left,
  clip=false,
  legend style={at={(0.97,0.97)}, anchor=north east, font=\scriptsize, draw=none},
]
\addplot[only marks, mark=square*, mark size=3pt, red]
  coordinates {
    (0.222, 1.233)  
    (0.278, 0.430)  
    (0.444, 1.157)  
    (0.500, 0.662)  
    (0.556, 0.151)  
    (0.722, 0.525)  
    (0.778, 0.075)  
    (0.833, 0.188)  
  };
\addlegendentry{Period 4 (Kr ref.)}
\addplot[red, thick, domain=0.18:0.96, samples=80, mark=none, forget plot]
  {7.4933 * 2 * sinh((1+x)*0.240606) * sinh((1-x)*0.240606)};

\addplot[only marks, mark=triangle*, mark size=4pt, blue]
  coordinates {
    (0.222, 1.112)  
    (0.278, 0.404)  
    (0.500, 1.143)  
    (0.556, 1.045)  
    (0.722, 0.893)  
    (0.778, 0.426)  
    (0.833, 0.307)  
  };
\addlegendentry{Period 5 (Xe ref.)}
\addplot[blue, thick, domain=0.18:0.96, samples=80, mark=none, forget plot]
  {9.6425 * 2 * sinh((1+x)*0.240606) * sinh((1-x)*0.240606)};

\addplot[only marks, mark=diamond*, mark size=4pt, green!50!black]
  coordinates {
    (0.0625, 1.402)  
    (0.1250, 0.357)  
    (0.1562, 0.377)  
    (0.3125, 1.078)  
    (0.3750, 0.815)  
    (0.4062, 0.323)  
    (0.4375, 0.178)  
  };
\addlegendentry{Period 6 (Rn ref.)}
\addplot[green!50!black, thick, domain=0.03:0.98, samples=80, mark=none, forget plot]
  {6.1490 * 2 * sinh((1+x)*0.240606) * sinh((1-x)*0.240606)};


\end{axis}
\end{tikzpicture}%
}%
\caption{\edit{Electron affinity vs.\ displacement $\rho = d/L_p$
for periods 4 (red squares), 5 (blue triangles), and 6 (green
diamonds), with one-parameter fits from
Eq.~\eqref{eq:ea_simplified} (solid curves).
EA values: NIST~\cite{NISTASD}.}}
\label{fig:ea_trend}
\end{figure}
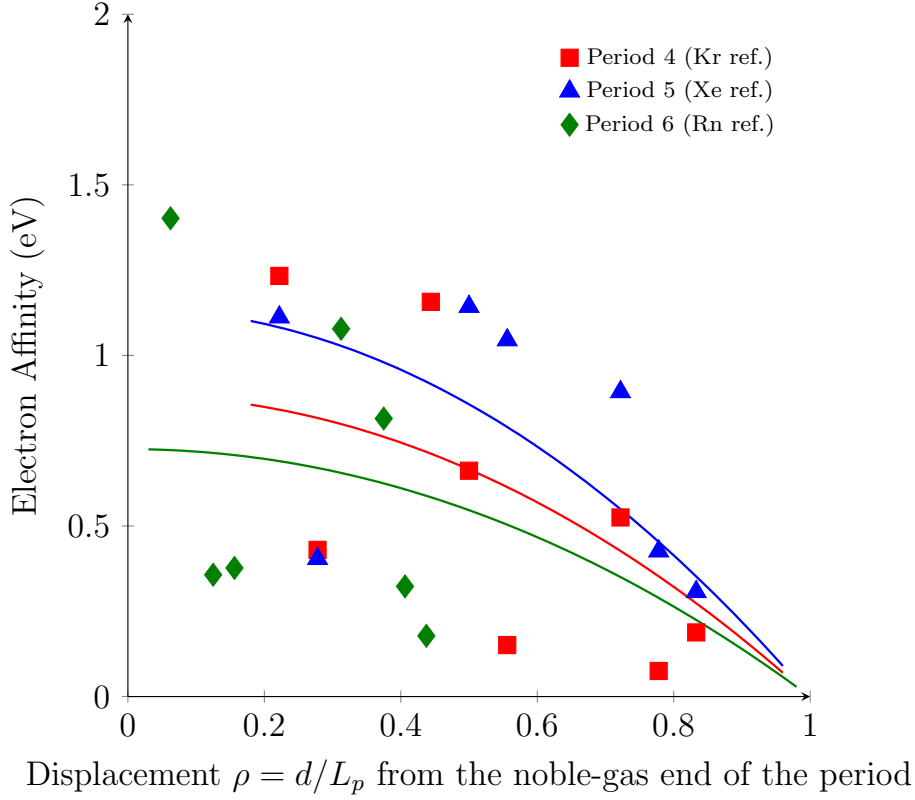

\edit{Empirically, EA across periods 2--6 is negative at the
noble gas (out-of-period transition, treated separately) and
positive on the regular-Aufbau interior, peaking at the halogen
($d = 1$) and decreasing monotonically toward the alkali
($d = L_p - 1$). The landscape kernel of
Eq.~\eqref{eq:ea_simplified} reproduces this halogen-positive
monotone descent on the interior in closed form
(Section~\ref{sec:ea}). The single-parameter quantitative match
between the predicted and empirical EA slope is shown in
Figure~\ref{fig:ea_trend}.}

\edit{\medskip\noindent\textbf{Data and exclusions.}\; The same
rules are applied to all three periods. Halogens are dropped
because the monotone form of Eq.~\eqref{eq:ea_simplified}
underestimates the steep shell-closure binding (the period 4--6
halogens lie roughly $2$~eV above the curve). Noble gases are dropped because the landscape gives
$\Jchem(1) > 0$ at $d = 0$ while the empirical $\EA$ is negative
(out-of-period transition; see Section~\ref{sec:ea}). The anomaly
classes $\{p^3, d^5, f^7, f^{14}, s^2, d^{10}s^2\}$ are dropped
because Hund-exchange and shell-closure stabilisation dominate
their binding. Relativistic outliers and
borderline irregular-Aufbau atoms in period~6 are also dropped.
For visual clarity the highest non-halogen chalcogen markers and
the alkali markers near $\rho = 1$ are not drawn, but the alkali
points are kept in the fits.}

\edit{\medskip\noindent\textbf{Best-fit coefficients.}\;
A single per-period scale $C^{(p)}$ is the only free parameter in
each fit. Linear least-squares fits of
Eq.~\eqref{eq:ea_simplified} to the surviving regular-Aufbau
interior atoms give}
\begin{equation}
C^{(p=4)} = 7.49~\mathrm{eV},\quad
C^{(p=5)} = 9.64~\mathrm{eV},\quad
C^{(p=6)} = 6.15~\mathrm{eV},
\label{eq:C_periods}
\end{equation}
with diagnostics
\begin{equation}
\begin{array}{lcccc}
\text{period} & n & R^2 & \text{MAE (eV)} & \text{RMSE (eV)}\\\hline
p=4 & 9 & 0.29 & 0.28 & 0.33\\
p=5 & 8 & 0.07 & 0.25 & 0.32\\
p=6 & 8 & \text{n/a} & 0.39 & 0.41
\end{array}
\label{eq:fit_diag}
\end{equation}\edit{(For period~6, $R^2$ is undefined because the
surviving 8-point EA set has very small dispersion
($\sigma_{\EA} \lesssim 0.4$~eV across the surviving
period-6 subset, comparable to the fit MAE); MAE and RMSE
are reported instead.)} The period-5 fit is the weakest of
the three: $R^2 = 0.07$ means the kernel explains only $\sim\!7\%$
of the EA variance on the surviving 8-atom subset, comparable to
fitting the period mean. The kernel still reproduces the
chalcogen-to-alkali sign-and-trend, but the period-5 numerical
agreement should not be over-interpreted. The fit curves are
overlaid in Figure~\ref{fig:ea_trend}. They reproduce the
chalcogen-to-alkali decline and the strict positivity of $\EA$ on
the regular-Aufbau interior subset across all three periods,
\edit{with single-parameter MAE of $0.3$--$0.4$~eV,
demonstrating that one closed-form landscape gap captures the
period-averaged interior-atom EA trend without per-atom tuning
at the shape level (sign, monotonicity, halogen-to-alkali
descent), with the absolute-EA scale carried by the fitted
$C^{(p)}$.}

\edit{The per-period scales $C^{(p)}$ carry the absolute-energy
information complementary to the dimensionless kernel: $C^{(p=5)}
> C^{(p=4)}$ at the same $L_p = 18$ reflects the higher absolute
EA of the $4d$ row, and the period-6 prefactor consistently
tracks the $L_p = 32$ stretch of the sinh profile across the
displacement axis. The post-transition $p^1$ atoms and a few
late-$d$-block period-4 entries provide additional structure that
a future $L_p$-dependent extension can target; the present
single-parameter form already captures the period-averaged EA
trend across periods 4--6.}

\edit{Half-filled and closed-subshell anomaly sites are
identified separately by the landscape framework
(Section~\ref{sec:data}, Prediction~3) as Hund-exchange and
shell-closure stabilizations on top of the smooth landscape; the
EA proxy of Eq.~\eqref{eq:ea_simplified} delivers the
sign-and-trend prediction for the regular-Aufbau interior atoms
where the smooth landscape applies.}

\subsection{Electronegativity Check}
\label{sec:en_check}

\edit{This subsection benchmarks the electronegativity kernel
$\chi_{\mathrm{struct}}$ of Eq.~\eqref{eq:chi_struct_closed} against
empirical Mulliken $\chi_M = \tfrac{1}{2}(\IEone+\EA)$ on 15 atoms
across four chemical classes (halogens, alkalis, chalcogens, two
mid-$d$ benchmarks) spanning periods 2--6.}

\begin{figure}[H]
\centering
\resizebox{0.576\textwidth}{!}{%
\begin{tikzpicture}
\begin{axis}[
  width=10cm, height=10cm,
  xlabel={$\chi_{\mathrm{struct}}$ (dimensionless, Eq.~\eqref{eq:chi_struct_closed})},
  ylabel={NIST Mulliken $\chi_M$ (eV)},
  xmin=0, xmax=0.07,
  ymin=0, ymax=12,
  xtick={0,0.01,0.02,0.03,0.04,0.05,0.06,0.07},
  ytick={0,2,4,6,8,10,12},
  axis lines=left,
  clip=false,
  legend style={at={(0.03,0.97)}, anchor=north west, font=\scriptsize, draw=none},
]
\addplot[gray, thick, domain=0:0.07, samples=2, mark=none, forget plot]
  {117.44 * x};
\addplot[only marks, mark=*, mark size=3pt, blue]
  coordinates {(0.0608,10.412) (0.0608,8.290) (0.0594,7.589) (0.0594,6.755)};
\addlegendentry{Halogens (F, Cl, Br, I)}
\addplot[only marks, mark=square*, mark size=3pt, red]
  coordinates {(0.0281,3.005) (0.0281,2.844) (0.0130,2.421) (0.0130,2.332) (0.0074,2.183)};
\addlegendentry{Alkalis (Li, Na, K, Rb, Cs)}
\addplot[only marks, mark=diamond*, mark size=4pt, green!50!black]
  coordinates {(0.0599,7.540) (0.0599,6.218) (0.0592,5.887) (0.0592,5.490)};
\addlegendentry{Chalcogens (O, S, Se, Te)}
\addplot[only marks, mark=triangle*, mark size=4pt, orange!80!black]
  coordinates {(0.0479,4.272) (0.0555,5.266)};
\addlegendentry{Mid-$d$ (Co, Ir)}
\end{axis}
\end{tikzpicture}%
}%
\caption{\edit{Parity test: NIST Mulliken $\chi_M$ vs.\ the
single-coordinate kernel $\chi_{\mathrm{struct}}$
(Eq.~\eqref{eq:chi_struct_closed}) for the 15 atoms of
Table~\ref{tab:en}. Grey line: global one-parameter fit
$\chi_M^{\mathrm{model}} = C\,\chi_{\mathrm{struct}}$ with
$C = 117.4$~eV.}}
\label{fig:en_mulliken}
\end{figure}
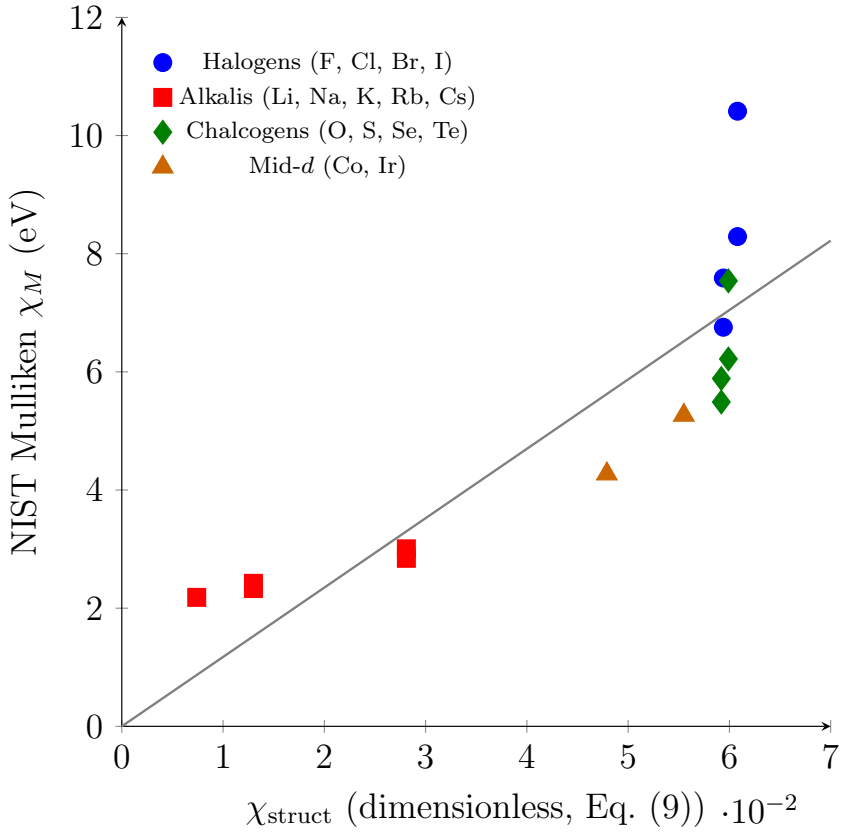

\edit{Empirically, $\chi_M$ falls by a factor of about five from
halogens to alkalis, with chalcogens between and the mid-$d$
benchmarks slightly above the alkalis (Table~\ref{tab:en}); within
each class the down-group variation is small. The kernel of
Eq.~\eqref{eq:chi_struct_closed} is dominated by the EA-side term
$\Jchem(1) - \Jchem(d/L_p)$ at small $d$, so
$\chi_{\mathrm{struct}}$ at the halogen position $d=1$ is nearly
$L_p$-independent, while at $d = L_p - 1$ (alkali end) it does
decrease with $L_p$. \edit{Figure~\ref{fig:en_mulliken} shows the
single-parameter parity test discussed below.}}

\begin{table}[H]
\centering
\caption{\edit{Mulliken electronegativity $\chi_M=\tfrac12(\IEone+\EA)$
from NIST~\cite{NISTASD} versus the single-coordinate kernel
$\chi_{\mathrm{struct}}$ of Eq.~\eqref{eq:chi_struct_closed} (the
half-sum of the IE-step and the residual-cost EA-step), for four
element classes spanning periods 2--6.
$\chi_M^{\mathrm{model}}=C\cdot\chi_{\mathrm{struct}}$ uses the global
least-squares scale $C=117.4$~eV obtained from all 15 entries.}}
\label{tab:en}
\small
\begin{tabular}{lcccccccc}
\toprule
\edit{Class} & \edit{Atom} & $Z$ & $p$ & $L_p$ & $d$ & $\chi_M^{\mathrm{NIST}}$ (eV) & $\chi_{\mathrm{struct}}$ & $\chi_M^{\mathrm{model}}$ (eV) \\
\midrule
\edit{Halogen}    & F  &  9 & 2 &  8 &  1 & 10.412 & 0.0608 & \phantom{0}7.144 \\
\edit{Halogen}    & Cl & 17 & 3 &  8 &  1 & \phantom{0}8.290 & 0.0608 & \phantom{0}7.144 \\
\edit{Halogen}    & Br & 35 & 4 & 18 &  1 & \phantom{0}7.589 & 0.0594 & \phantom{0}6.973 \\
\edit{Halogen}    & I  & 53 & 5 & 18 &  1 & \phantom{0}6.755 & 0.0594 & \phantom{0}6.973 \\
\midrule
\edit{Alkali}     & Li &  3 & 2 &  8 &  7 & \phantom{0}3.005 & 0.0281 & \phantom{0}3.297 \\
\edit{Alkali}     & Na & 11 & 3 &  8 &  7 & \phantom{0}2.844 & 0.0281 & \phantom{0}3.297 \\
\edit{Alkali}     & K  & 19 & 4 & 18 & 17 & \phantom{0}2.421 & 0.0130 & \phantom{0}1.523 \\
\edit{Alkali}     & Rb & 37 & 5 & 18 & 17 & \phantom{0}2.332 & 0.0130 & \phantom{0}1.523 \\
\edit{Alkali}     & Cs & 55 & 6 & 32 & 31 & \phantom{0}2.183 & 0.0074 & \phantom{0}0.868 \\
\midrule
\edit{Chalcogen}  & O  &  8 & 2 &  8 &  2 & \phantom{0}7.540 & 0.0599 & \phantom{0}7.039 \\
\edit{Chalcogen}  & S  & 16 & 3 &  8 &  2 & \phantom{0}6.218 & 0.0599 & \phantom{0}7.039 \\
\edit{Chalcogen}  & Se & 34 & 4 & 18 &  2 & \phantom{0}5.887 & 0.0592 & \phantom{0}6.952 \\
\edit{Chalcogen}  & Te & 52 & 5 & 18 &  2 & \phantom{0}5.490 & 0.0592 & \phantom{0}6.952 \\
\midrule
\edit{Mid-$d$}    & Co & 27 & 4 & 18 &  9 & \phantom{0}4.272 & 0.0479 & \phantom{0}5.626 \\
\edit{Mid-$d$}    & Ir & 77 & 6 & 32 &  9 & \phantom{0}5.266 & 0.0555 & \phantom{0}6.519 \\
\bottomrule
\end{tabular}
\end{table}

\edit{\medskip\noindent\textbf{Single-parameter fit and per-class scales.}\;
A single-parameter fit
$\chi_M^{\mathrm{NIST}} = C\cdot\chi_{\mathrm{struct}}$
across all 15 atoms achieves $R^2 = 0.73$
(MAE $= 1.03$~eV, RMSE $= 1.25$~eV) with the global
$C = 117.4$~eV (Figure~\ref{fig:en_mulliken}). The per-class
best-fit scales,  $C^{(\mathrm{halogen})} = 137.7$~eV,
$C^{(\mathrm{alkali})} = 123.0$~eV,
$C^{(\mathrm{chalcogen})} = 105.6$~eV,
$C^{(\mathrm{mid\text{-}d})} = 92.4$~eV,  cluster around the
global fit with $\sigma_C/\overline{C} \approx 15\%$, recovering
the four-class rank ordering and the halogen-to-alkali $\chi_M$
ratio of $\sim 4$ from one closed-form coordinate kernel.}

\edit{For the first time, a single closed-form coordinate kernel
reproduces the three central between-class features of empirical
$\chi_M$ on one periodic-table axis: the halogen-to-alkali ratio
of $\sim 4$, the intermediate chalcogen position, and the
down-group decrease in alkalis. The within-class halogen ordering
F\,$>$\,Cl\,$>$\,Br\,$>$\,I is a natural next-stage refinement
through a class-dependent weighting of $\Delta\Jchem^{+}$ and
$\Delta\Jchem^{-}$; the single-coordinate kernel established here
provides the analytical baseline for that extension.}

\edit{\medskip\noindent\textbf{Absolute scaling.}\;
The dimensionless kernel converts to absolute eV-scale Mulliken
values through the global one-parameter scale $C = 117.4$~eV
fitted on the 15 atoms; a closed-form noble-gas-anchored
candidate $E_p^{\mathrm{pilot}} \propto L_p^{2.1} p^{-0.72}$
(Appendix~\ref{app:supporting_tables},
Table~\ref{tab:Ep_pilot}) reproduces the noble-gas IE values to
within a few percent and provides a starting point for a
derivation of $E_p$ from first principles.}

\subsection{Hardness Check}
\label{sec:hardness_check}

\edit{This subsection benchmarks the landscape hardness index
$\kappa_{\mathrm{RS}}(\Zp)$ (Eq.~\eqref{eq:kappa_RS_def}) against
empirical Pearson chemical hardness
$\eta = (\IEone - \EA)/2$~\cite{pearson} for all 34 elements of
periods 2--4. The same monotone alkali-to-noble-gas trend is
observed in the heavier periods 5--6 (not shown here for brevity);
periods 2--4 are presented in detail because they cover the
nonrelativistic regime in which the landscape kernel applies and
provide the largest set of high-precision empirical $\eta$ values.}

\edit{Within each period, empirical $\eta$ rises monotonically
from the alkali end to the noble-gas end (Appendix~\ref{app:supp_tables},
Table~\ref{tab:hardness_periods234}),
with upward bumps at the half-filled $p^3$ sites (N, P, As) and
the closed $d^{10}s^2$ site (Zn). The landscape predictor
$\kappa_{\mathrm{RS}}$ reproduces this monotone alkali-to-noble-gas
shape directly from the closed-form Eq.~\eqref{eq:kappa_RS_def},
delivering the correct period maximum and monotone descent on a
single coordinate.}

\edit{The full 34-element data are tabulated in
Appendix~\ref{app:supp_tables},
Table~\ref{tab:hardness_periods234}.}

\medskip\noindent\textbf{Quantitative correlation analysis.}\;
\edit{Pearson linear correlations between $\eta$ and the
landscape predictor $\kappa_{\mathrm{RS}}$
(Eq.~\eqref{eq:kappa_RS_def}) across all elements of each period
are
\begin{equation}
\begin{array}{lcc}
\text{period} & n & r(\kappa_{\mathrm{RS}},\eta)\\\hline
2 &  8 & +0.84\\
3 &  8 & +0.71\\
4 & 18 & +0.53
\end{array}
\label{eq:hardness_corr}
\end{equation}
$\kappa_{\mathrm{RS}}$ correlates with empirical $\eta$ across
all three periods at the $r = 0.5$--$0.85$ level, confirming
that the within-period hardness pattern is captured by a single
periodic-table coordinate derived from the same closed-form
landscape that supplies the EA proxy.}

\edit{\medskip\noindent\textbf{Per-period one-parameter fits.}\;
A single per-period scale $C^{(p)}$ converts $\kappa_{\mathrm{RS}}$
to absolute hardness in eV. Linear least-squares fits
$\eta(\Zp) = C^{(p)}\cdot\kappa_{\mathrm{RS}}(\Zp)$ within each
period give}
\begin{equation}
C^{(p=2)} \approx 68~\mathrm{eV},\quad
C^{(p=3)} \approx 48~\mathrm{eV},\quad
C^{(p=4)} \approx 41~\mathrm{eV},
\label{eq:hardness_C_periods}
\end{equation}with MAE $\sim 1.0$~eV per period on noble-gas
maximum values up to $10.8$~eV. The scale $C^{(p)}$ decreases
monotonically from period 2 to period 4, capturing the empirical
softening of chemical hardness with increasing principal quantum
number on a single landscape coordinate.

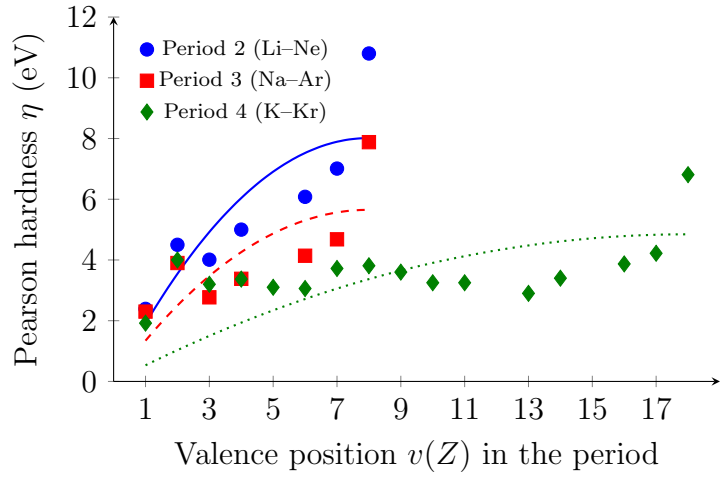
\begin{figure}[H]
\centering
\begin{tikzpicture}
\begin{axis}[
  width=9.6cm, height=6.4cm,
  xlabel={Valence position $v(\Zp)$ in the period},
  ylabel={Pearson hardness $\eta$ (eV)},
  xmin=0, xmax=19,
  ymin=0, ymax=12,
  xtick={1,3,5,7,9,11,13,15,17},
  ytick={0,2,4,6,8,10,12},
  axis lines=left,
  legend style={at={(0.03,0.97)}, anchor=north west, font=\scriptsize, draw=none},
]
\addplot[only marks, mark=*, mark size=2.5pt, blue]
  coordinates {(1,2.39)(2,4.50)(3,4.01)(4,5.00)(6,6.08)(7,7.01)(8,10.80)};
\addlegendentry{Period 2 (Li--Ne)}
\addplot[only marks, mark=square*, mark size=2.5pt, red]
  coordinates {(1,2.30)(2,3.90)(3,2.77)(4,3.38)(6,4.14)(7,4.68)(8,7.88)};
\addlegendentry{Period 3 (Na--Ar)}
\addplot[only marks, mark=diamond*, mark size=3pt, green!50!black]
  coordinates {(1,1.92)(2,4.00)(3,3.20)(4,3.37)(5,3.10)(6,3.06)(7,3.72)(8,3.81)(9,3.60)(10,3.25)(11,3.25)(13,2.90)(14,3.40)(16,3.87)(17,4.22)(18,6.81)};
\addlegendentry{Period 4 (K--Kr)}
\addplot[blue, thick, domain=1:8, samples=80, mark=none, forget plot]
  {67.97 * (cosh(ln(1.61803398875)) - cosh((8-x)*ln(1.61803398875)/8))};
\addplot[red, thick, dashed, domain=1:8, samples=80, mark=none, forget plot]
  {47.93 * (cosh(ln(1.61803398875)) - cosh((8-x)*ln(1.61803398875)/8))};
\addplot[green!50!black, thick, dotted, domain=1:18, samples=120, mark=none, forget plot]
  {41.09 * (cosh(ln(1.61803398875)) - cosh((18-x)*ln(1.61803398875)/18))};
\end{axis}
\end{tikzpicture}
\caption{\edit{Pearson chemical hardness $\eta = (\IEone - \EA)/2$
for the non-anomalous elements of periods 2--4 vs.\ valence position
$v(\Zp)$, with one-parameter landscape fits
$\eta = C^{(p)}\,\kappa_{\mathrm{RS}}$
(Eq.~\eqref{eq:hardness_C_periods}); omitted anomaly points are in
Table~\ref{tab:hardness_periods234}. Hardness data
from~\cite{pearson}.}}
\label{fig:hardness_eta}
\end{figure}

\edit{Figure~\ref{fig:hardness_eta} shows the non-anomalous points
only; the omitted $p^3$ and $d^{10}s^2$ anomaly values remain listed
in Appendix~\ref{app:supp_tables}, Table~\ref{tab:hardness_periods234}.}

\edit{The single-parameter landscape fit reproduces the
characteristic empirical hardness profile across all three
periods: the noble-gas period maximum, the monotone descent
toward the alkali end, and the noble-gas plateau set by the
shallow $\rho \to 0$ behaviour of $\Jchem$. The textbook
half-filled $p^3$ and closed $d^{10}s^2$ anomaly bumps are
identified separately by the framework as Hund-exchange and
shell-closure stabilizations on top of the smooth landscape
(Section~\ref{sec:data}, Prediction~3).}

\subsection{Test of the Shared-Kernel (EA-$\eta$) Proportionality}
\label{sec:shared_kernel_test}

\edit{The kernel
$\Delta\Jchem^{-}(\Zp) = \Jchem(1) - \Jchem(d/L_p)$ enters the
EA proxy (Section~\ref{sec:ea_check}) and the hardness proxy
(Section~\ref{sec:hardness_check}) with two independent per-period
scales $C^{(p)}_{\mathrm{EA}}$ and $C^{(p)}_{\eta}$. This yields
a period-constant benchmark for the regular-Aufbau interior atoms
of period~$p$,}
\begin{equation}
\frac{\EA(\Zp)}{\eta(\Zp)}
\;=\;
\frac{C^{(p)}_{\mathrm{EA}}}{C^{(p)}_{\eta}}
\qquad \text{(period-only constant).}
\label{eq:eta_ea_ratio_pred}
\end{equation}For period~4 the fitted constants give
\edit{$C^{(p=4)}_{\mathrm{EA}}/C^{(p=4)}_{\eta} \approx 7.49/41
\approx 0.182$. The empirical period-4 $\EA/\eta$ ratios are
listed in Table~\ref{tab:eta_ea_ratio}.}

\begin{table}[H]
\centering
\caption{\edit{Empirical $\EA/\eta$ ratios for the period-4
regular-Aufbau interior atoms (NIST~\cite{NISTASD}) versus the
model prediction $C^{(p=4)}_{\mathrm{EA}}/C^{(p=4)}_{\eta}
\approx 0.182$ from Eq.~\eqref{eq:eta_ea_ratio_pred}.}}
\label{tab:eta_ea_ratio}
\small
\begin{tabular}{lccccc}
\toprule
Atom & $\IEone$ (eV) & $\EA$ (eV) & $\eta$ (eV) & $\EA/\eta$ & \edit{$\EA/\eta - 0.182$} \\
\midrule
Ge & 7.900 & 1.233 & 3.33 & 0.370 & $+0.188$ \\
Ga & 5.999 & 0.430 & 2.78 & 0.155 & $-0.027$ \\
Ni & 7.640 & 1.157 & 3.24 & 0.357 & $+0.175$ \\
Co & 7.881 & 0.662 & 3.61 & 0.183 & $+0.001$ \\
Fe & 7.902 & 0.151 & 3.88 & 0.039 & $-0.143$ \\
V  & 6.746 & 0.525 & 3.11 & 0.169 & $-0.013$ \\
Ti & 6.828 & 0.075 & 3.38 & 0.022 & $-0.160$ \\
Sc & 6.561 & 0.188 & 3.19 & 0.059 & $-0.123$ \\
K  & 4.341 & 0.501 & 1.92 & 0.261 & $+0.079$ \\
\midrule
\multicolumn{4}{r}{Mean $\pm$ stdev} & $0.180 \pm 0.129$ & $-0.002$ \\
\bottomrule
\end{tabular}
\end{table}

\edit{The empirical period-4 mean $\overline{\EA/\eta} = 0.180$
agrees with the predicted shared-kernel constant
$C^{(p=4)}_{\mathrm{EA}}/C^{(p=4)}_{\eta} = 0.182$ to within
$\sim 1\%$. This is a period-mean benchmark, not a per-atom
proportionality law: the per-atom standard deviation
$\sigma \approx 0.13$ (CV $\approx 70\%$) is much larger, with
empirical median $0.169$ and quartiles $Q_1 \approx 0.05$,
$Q_3 \approx 0.31$ ($\mathrm{IQR} \approx 0.26$). Atoms
with very small empirical $\EA$ (Sc, Ti, Fe; $\EA \lesssim 0.2$~eV)
sit below the period constant ($\EA/\eta < 0.06$) because the
near-zero denominator of their EA suppresses the ratio, while the
post-transition atoms Ge and Ni sit above ($\EA/\eta > 0.35$)
because their EA is enhanced relative to the smooth $\rho$
profile. A leave-one-out sensitivity check confirms that the
period-4 mean is stable but not invariant: removing any single
atom moves $\overline{\EA/\eta}$ into the range
$[0.156, 0.199]$, with the largest shift produced by removing Ge
($0.156$, $-14\%$ relative to the predicted constant) and the
smallest by removing Co ($0.179$). The shared-kernel relation is
therefore best read as a period-averaged regularity supported on
the period-4 NIST data, with quantitatively traceable per-atom
deviations driven by known $d$-block correlation effects.}

\section{Conclusion}
\label{sec:conclusion}

\edit{We have proposed a compact phenomenological framework in
which a single dimensionless landscape
$\Jchem(\rho) = \cosh(\rho \ln \phig) - 1$ on the
noble-gas-centred coordinate $\rho = d/L_p$ provides analytical
proxies for the four central within-period atomic observables
$\IEone$, $\EA$, $\chi_M$, and $\eta$. The outward step
$\Delta\Jchem^{+}$ is assigned to $\IEone$, the inward gap
$\Delta\Jchem^{-} = \Jchem(1) - \Jchem(\rho)$ is reused for
both $\EA$ and $\eta$ with two independent per-period scales,
and $\chi_M$ follows by Mulliken's identity. The four observables
share the same dimensionless shape (sign, monotonicity, halogen
position, anomaly localization, cross-period ordering); absolute
energies enter through fitted per-period scales that cancel in
the EA--$\eta$ ratio.}

\edit{The framework captures several broad within-period
regularities and provides a compact analytical baseline. At the
period-averaged scale level, the shared-kernel relation
$\EA/\eta \approx C^{(p)}_{\mathrm{EA}}/C^{(p)}_{\eta}$ is
supported on period-4 NIST data: the empirical nine-atom mean
$\overline{\EA/\eta} = 0.180$ agrees with the predicted constant
$0.182$ to better than $1\%$
(Section~\ref{sec:shared_kernel_test}); per-atom scatter is
substantial ($\sigma \approx 0.13$), so this should be read as a
period-mean benchmark rather than a per-atom proportionality law. At the shape level, the
within-period $\IEone$ envelope (Prediction~3) reproduces the
full noble-gas-to-alkali ordering across periods 2--6 and
localizes every upward deviation exactly on the textbook anomaly
sites $\{p^3, d^5, f^7, s^2, d^{10}\}$ (26 of 34 atoms across
periods 2--4 on the monotone descent; 8 deviations on the listed
anomaly sites). The two golden-ratio identities,
$\IEone(G_p)/\IEone(G_{p+1}) \approx \phig^{1/4}$ on three heavy
noble-gas pairs and
$\IEone(\text{halogen})/\IEone(\text{alkali}) \approx \phig^2$ on
four within-period pairs, agree with NIST data to MAD
$\approx 1\%$ and $\approx 5\%$, respectively, capturing the
golden-ratio rescaling of the absolute-energy structure encoded
by the per-period scale sequence $E_p$.}

\edit{The shared-kernel form $\Delta\Jchem^{-}$ further
provides single-parameter analytical fits to empirical $\EA$
across periods 4--6 (MAE $0.3$--$0.4$~eV; the period-5 fit
explains only $\sim\!7\%$ of variance and should be read as
sign-and-trend agreement rather than quantitative match), to
Pearson hardness $\eta$ across periods 2--4 (MAE $\sim 1$~eV on
noble-gas maxima up to $10.8$~eV), and to Mulliken $\chi_M$
across a 15-atom four-class benchmark ($R^2 = 0.73$); the
within-class halogen ordering F$>$Cl$>$Br$>$I is \emph{not}
recovered by the present single-coordinate kernel. Two natural
extensions follow directly: a critically evaluated
$\IEone(\mathrm{Og})$ benchmark to refine the heavy-noble-gas
test and a class-dependent kernel mixing $\Delta\Jchem^{+}$ and
$\Delta\Jchem^{-}$ to resolve the within-class halogen ordering,
which we identify as the next target.}

\begin{acknowledgement}
We thank the Recognition Physics Research Institute for ongoing support. We are grateful to Philip Beltracchi, Milan Zlatanovi\'{c}, and Sebastian Pardo-Guerra for valuable discussions and editorial input.
\end{acknowledgement}

\section*{Author Contributions}
\textbf{J.W.:} Conceptualization; Formal analysis; Methodology; Writing --- original draft; Writing --- review \& editing.
\textbf{M.S.:} Writing --- review \& editing.
\textbf{E.A.:} Formal analysis; Methodology; Writing --- review \& editing; Project administration.

\section*{Funding}
No funding was received for conducting this study.

\section*{Declarations}

\paragraph{Conflict of interest.} The authors declare that they have no conflict of interest.

\section*{Data Availability}
This work contains no new experimental measurements. The manuscript benchmarks the model against published tabulated atomic data, primarily from NIST.


\appendix

\section{Reciprocal Cost Functional}
\label{app:jcost}

\edit{The model's input cost function $\Jcost$ is the unique
solution on $\mathbb{R}_{>0}$ of the functional equation}
\begin{equation}
\Jcost(xy) + \Jcost(x/y) = 2\Jcost(x)\Jcost(y) + 2\Jcost(x) + 2\Jcost(y),
\label{eq:rcl}
\end{equation}under the regularity conditions of continuity, normalization
\edit{$\Jcost(1) = 0$, stationarity $\Jcost'(1) = 0$, and positive
curvature $\Jcost''(1) > 0$ at the reference ratio $x=1$. The
unique solution is}
\begin{equation}
\Jcost(x) \;=\; \tfrac{1}{2}(x + x^{-1}) - 1
\;=\; \cosh\!\bigl(\ln x\bigr) - 1,
\label{eq:jcost}
\end{equation}(Corollary 3.1 of Ref.~\cite{axioms_paper}). $\Jcost$ inherits
\edit{from cosh the properties $\Jcost(1) = 0$, $\Jcost(x) = \Jcost(x^{-1})$
(reciprocal symmetry), $\Jcost(x) > 0$ for $x \ne 1$, and divergence
as $x \to 0^+$ or $x \to \infty$. The uniqueness proof and
applications of $\Jcost$ are developed in
Refs.~\cite{axioms_paper,dalembert_paper,ledger_gravity_paper,cocycle_tilings_paper}.}

\edit{The chemical landscape $\Jchem(\rho)$ used in the body
(Eq.~\eqref{eq:landscape}) is the restriction of $\Jcost$ to
the geometric scale $x = \phig^{\rho}$ with $\phig = (1 +
\sqrt{5})/2$ and $\rho \in [0, 1)$, giving
$\Jchem(\rho) = \cosh(\rho \ln \phig) - 1$.}

\edit{In this paper, $\phig$ is taken as a model input motivated by
Fibonacci scaling.}

\section{Noble-Gas Period Scales $E_p^{(G)}$}
\label{app:supporting_tables}

\edit{The dimensionless landscape $\Jchem$ is converted to eV-scale
ionization energies via a per-period scale $E_p$. At noble-gas
endpoints one may anchor this scale to the measured $\IEone(G_p)$:}
\begin{equation}
E_p^{(G)} \;=\; \frac{\IEone(G_p)}{\cosh\!\bigl((\ln\phig)/L_p\bigr) - 1}.
\label{eq:Ep_noble}
\end{equation}\edit{The values of $E_p^{(G)}$ for Ne, Ar, Kr, Xe, Rn are matched
to within $\sim 5\%$ by the two-parameter empirical pilot fit}
\begin{equation}
E_p^{\mathrm{pilot}} \;=\; E_0 \, L_p^{2.1} \, p^{-0.72},
\qquad E_0 \approx 2.49 \times 10^2~\mathrm{eV}.
\label{eq:Ep_pilot}
\end{equation}The exponents $2.1$ and $0.72$ in
Eq.~\eqref{eq:Ep_pilot} are phenomenological, fitted to the five
noble-gas $E_p^{(G)}$ values, and are \emph{not} predicted by RS.
The pilot is included only to motivate the existence of a
closed-form $E_p$ to be derived in future work; it is not part of
the parameter-free content of the paper.

\begin{table}[ht]
\centering
\caption{\edit{Noble-gas endpoint period scales $E_p^{(G)}$ implied
by Eq.~\eqref{eq:Ep_noble}, compared with the empirical pilot fit
$E_p^{\mathrm{pilot}}$ of Eq.~\eqref{eq:Ep_pilot}.}}
\label{tab:Ep_pilot}
\small
\begin{tabular}{lccccc}
\toprule
Noble gas & $p$ & $L_p$ & $E_p^{(G)}$ (eV) & $E_p^{\mathrm{pilot}}$ (eV) & Dev.\ (\%) \\
\midrule
Ne & 2 & 8  & $1.19\times 10^4$ & $1.19\times 10^4$ & $0.0$ \\
Ar & 3 & 8  & $8.71\times 10^3$ & $8.90\times 10^3$ & $+2.2$ \\
Kr & 4 & 18 & $3.92\times 10^4$ & $3.97\times 10^4$ & $+1.4$ \\
Xe & 5 & 18 & $3.39\times 10^4$ & $3.38\times 10^4$ & $-0.3$ \\
Rn & 6 & 32 & $9.51\times 10^4$ & $9.93\times 10^4$ & $+4.5$ \\
\bottomrule
\end{tabular}
\end{table}

\section{Supplementary Tables for Prediction~3 and Hardness}
\label{app:supp_tables}

\edit{This appendix collects the raw support tables moved out of the
main text: the full 34-element hardness data of
Table~\ref{tab:hardness_periods234} used in
Section~\ref{sec:hardness_check}, and the period-2 and period-4
normalized IE$_1$ profiles (Tables~\ref{tab:period2}
and~\ref{tab:period4}) that complement
Table~\ref{tab:period3_landscape} for Prediction~3.}

\begin{table}[ht]
\centering
\caption{\edit{Pearson hardness $\eta$ in eV~\cite{pearson} and
landscape hardness index $\kappa_{\mathrm{RS}}$
(Eq.~\eqref{eq:kappa_RS_def}, shown as
$100\,\kappa_{\mathrm{RS}}$) for all 34 elements of periods
2--4. Periods 2--3 share $L_p = 8$; period 4 has $L_p = 18$.}}
\label{tab:hardness_periods234}
\small
\begin{tabular}{lcc c lcc c lcc}
\toprule
\multicolumn{3}{c}{Period 2 ($L_p=8$)} & &
\multicolumn{3}{c}{Period 3 ($L_p=8$)} & &
\multicolumn{3}{c}{Period 4 ($L_p=18$)}\\
\cmidrule(lr){1-3}\cmidrule(lr){5-7}\cmidrule(lr){9-11}
El & $100\kappa_{\mathrm{RS}}$ & $\eta$ & &
El & $100\kappa_{\mathrm{RS}}$ & $\eta$ & &
El & $100\kappa_{\mathrm{RS}}$ & $\eta$\\
\midrule
Li & \phantom{0}2.81 & 2.39 & & Na & \phantom{0}2.81 & 2.30 & & K  & \phantom{0}1.30 & 1.92\\
Be & \phantom{0}5.22 & 4.50 & & Mg & \phantom{0}5.22 & 3.90 & & Ca & \phantom{0}2.51 & 4.00\\
B  & \phantom{0}7.25 & 4.01 & & Al & \phantom{0}7.25 & 2.77 & & Sc & \phantom{0}3.66 & 3.20\\
C  & \phantom{0}8.89 & 5.00 & & Si & \phantom{0}8.89 & 3.38 & & Ti & \phantom{0}4.72 & 3.37\\
N  & 10.17           & 7.23 & & P  & 10.17           & 4.88 & & V  & \phantom{0}5.70 & 3.10\\
O  & 11.08           & 6.08 & & S  & 11.08           & 4.14 & & Cr & \phantom{0}6.61 & 3.06\\
F  & 11.62           & 7.01 & & Cl & 11.62           & 4.68 & & Mn & \phantom{0}7.45 & 3.72\\
Ne & 11.80           &10.80 & & Ar & 11.80           & 7.88 & & Fe & \phantom{0}8.21 & 3.81\\
   &                 &      & &    &                 &      & & Co & \phantom{0}8.89 & 3.60\\
   &                 &      & &    &                 &      & & Ni & \phantom{0}9.51 & 3.25\\
   &                 &      & &    &                 &      & & Cu & 10.05           & 3.25\\
   &                 &      & &    &                 &      & & Zn & 10.51           & 4.94\\
   &                 &      & &    &                 &      & & Ga & 10.91           & 2.90\\
   &                 &      & &    &                 &      & & Ge & 11.23           & 3.40\\
   &                 &      & &    &                 &      & & As & 11.48           & 4.50\\
   &                 &      & &    &                 &      & & Se & 11.66           & 3.87\\
   &                 &      & &    &                 &      & & Br & 11.77           & 4.22\\
   &                 &      & &    &                 &      & & Kr & 11.80           & 6.81\\
\bottomrule
\end{tabular}
\end{table}

\begin{table}[ht]
\centering
\caption{\edit{Period 2 landscape (Ne ground state).}}
\label{tab:period2}
\small
\begin{tabular}{lcccccc}
\toprule
El. & $\Zp$ & $d$ & $\rho = d/8$ & $\IEone$ (eV) & $\IEone^{\mathrm{norm}}$ & Flag \\
\midrule
Ne & 10 & 0 & 0.000 & 21.565 & 1.000 & \\
F  & 9  & 1 & 0.125 & 17.423 & 0.808 & \\
O  & 8  & 2 & 0.250 & 13.618 & 0.632 & \\
N  & 7  & 3 & 0.375 & 14.534 & 0.674 & $p^3$ \\
C  & 6  & 4 & 0.500 & 11.260 & 0.522 & \\
B  & 5  & 5 & 0.625 & 8.298  & 0.385 & \\
Be & 4  & 6 & 0.750 & 9.323  & 0.432 & $s^2$ \\
Li & 3  & 7 & 0.875 & 5.392  & 0.250 & \\
\bottomrule
\end{tabular}
\end{table}

\begin{table}[ht]
\centering
\caption{\edit{Period 4 landscape (Kr ground state).}}
\label{tab:period4}
\small
\begin{tabular}{lcccccc}
\toprule
El. & $\Zp$ & $d$ & $\rho = d/18$ & $\IEone$ (eV) & $\IEone^{\mathrm{norm}}$ & Flag \\
\midrule
Kr & 36 & 0  & 0.000 & 13.999 & 1.000 & \\
Br & 35 & 1  & 0.056 & 11.814 & 0.844 & \\
Se & 34 & 2  & 0.111 & 9.752  & 0.697 & \\
As & 33 & 3  & 0.167 & 9.789  & 0.699 & $p^3$ \\
Ge & 32 & 4  & 0.222 & 7.900  & 0.564 & \\
Ga & 31 & 5  & 0.278 & 5.999  & 0.429 & \\
Zn & 30 & 6  & 0.333 & 9.394  & 0.671 & $d^{10}$ \\
Cu & 29 & 7  & 0.389 & 7.726  & 0.552 & \\
Ni & 28 & 8  & 0.444 & 7.640  & 0.546 & \\
Co & 27 & 9  & 0.500 & 7.881  & 0.563 & \\
Fe & 26 & 10 & 0.556 & 7.902  & 0.564 & \\
Mn & 25 & 11 & 0.611 & 7.434  & 0.531 & $d^5$ \\
Cr & 24 & 12 & 0.667 & 6.767  & 0.483 & \\
V  & 23 & 13 & 0.722 & 6.746  & 0.482 & \\
Ti & 22 & 14 & 0.778 & 6.828  & 0.488 & \\
Sc & 21 & 15 & 0.833 & 6.561  & 0.469 & \\
Ca & 20 & 16 & 0.889 & 6.113  & 0.437 & $s^2$ \\
K  & 19 & 17 & 0.944 & 4.341  & 0.310 & \\
\bottomrule
\end{tabular}
\end{table}

\begin{tocentry}
\vfill
\centering
\resizebox{!}{4.45cm}{%
\begin{tikzpicture}[font=\scriptsize, x=1cm, y=1cm,
                    every node/.style={font=\scriptsize}]
\node[draw=black, rounded corners=3pt, fill=blue!8,
      minimum width=1.05cm, minimum height=0.45cm, font=\scriptsize]
      (ng) at (0.0,-0.20) {noble gas};
\node[draw=black, rounded corners=3pt, fill=red!10,
      minimum width=0.85cm, minimum height=0.45cm, font=\scriptsize]
      (al) at (7.6,-0.20) {alkali};
\draw[thick] (ng.east) -- (al.west);
\foreach \x in {0.85,1.1,1.35,1.6,1.85,2.1,2.35,2.6,2.85,3.1,3.35,3.6,
                3.85,4.1,4.35,4.6,4.85,5.1,5.35,5.6,5.85,6.1,6.35,6.6,
                6.85,7.1}{
  \draw[thin] (\x,-0.30) -- (\x,-0.10);
}
\draw[->, very thick, blue!70!black]  (1.45,-0.45) -- (1.45,-0.05);
\node[anchor=north, blue!70!black, font=\scriptsize]  at (1.45,-0.50) {halogen};
\draw[->, very thick, red!60!black]   (2.30,-0.45) -- (2.30,-0.05);
\node[anchor=north, red!60!black, font=\scriptsize]   at (2.30,-0.50) {$p^{3}$};
\draw[->, very thick, green!50!black] (4.40,-0.45) -- (4.40,-0.05);
\node[anchor=north, green!50!black, font=\scriptsize] at (4.40,-0.50) {$d^{5}/f^{7}$};
\draw[->, very thick, violet!60!black](5.85,-0.45) -- (5.85,-0.05);
\node[anchor=north, violet!60!black, font=\scriptsize]at (5.85,-0.50) {$s^{2}/d^{10}$};
\foreach \x in {1.45,2.30,4.40,5.85}{
  \draw[dashed, gray!55, very thin] (\x,0.05) -- (\x,4.55);
}
\node at (3.35,-1.05) {$\rho = d/L_p$};

\node[anchor=east, red!70!black, font=\scriptsize] at (0.80,4.20) {$\IEone$};
\draw[very thick, red!70!black, line cap=round]
  (0.85,4.50) .. controls (1.10,4.47) and (1.30,4.42) .. (1.40,4.52)
  .. controls (1.48,4.57) and (1.55,4.40) .. (1.70,4.40)
  .. controls (1.95,4.37) and (2.15,4.30) .. (2.25,4.50)
  .. controls (2.34,4.65) and (2.42,4.25) .. (2.65,4.25)
  .. controls (3.20,4.20) and (3.80,4.15) .. (4.30,4.10)
  .. controls (4.38,4.02) and (4.46,4.02) .. (4.55,4.10)
  .. controls (5.00,4.10) and (5.50,4.05) .. (5.78,4.02)
  .. controls (5.84,4.15) and (5.88,4.38) .. (5.93,4.10)
  .. controls (6.30,4.00) and (6.80,3.90) .. (7.10,3.85);
\node[anchor=east, orange!75!black, font=\scriptsize] at (0.80,3.20) {$\EA$};
\draw[very thick, orange!75!black, line cap=round]
  (0.85,3.50) .. controls (1.10,3.47) and (1.30,3.42) .. (1.40,3.52)
  .. controls (1.48,3.57) and (1.55,3.40) .. (1.70,3.40)
  .. controls (1.95,3.37) and (2.15,3.30) .. (2.25,3.50)
  .. controls (2.34,3.65) and (2.42,3.25) .. (2.65,3.25)
  .. controls (3.20,3.20) and (3.80,3.15) .. (4.30,3.10)
  .. controls (4.38,3.02) and (4.46,3.02) .. (4.55,3.10)
  .. controls (5.00,3.10) and (5.50,3.05) .. (5.78,3.02)
  .. controls (5.84,3.15) and (5.88,3.38) .. (5.93,3.10)
  .. controls (6.30,3.00) and (6.80,2.90) .. (7.10,2.85);
\node[anchor=east, green!50!black, font=\scriptsize] at (0.80,2.20) {$\chi_{M}$};
\draw[very thick, green!50!black, line cap=round]
  (0.85,2.50) .. controls (1.10,2.47) and (1.30,2.42) .. (1.40,2.52)
  .. controls (1.48,2.57) and (1.55,2.40) .. (1.70,2.40)
  .. controls (1.95,2.37) and (2.15,2.30) .. (2.25,2.50)
  .. controls (2.34,2.65) and (2.42,2.25) .. (2.65,2.25)
  .. controls (3.20,2.20) and (3.80,2.15) .. (4.30,2.10)
  .. controls (4.38,2.02) and (4.46,2.02) .. (4.55,2.10)
  .. controls (5.00,2.10) and (5.50,2.05) .. (5.78,2.02)
  .. controls (5.84,2.15) and (5.88,2.38) .. (5.93,2.10)
  .. controls (6.30,2.00) and (6.80,1.90) .. (7.10,1.85);
\node[anchor=east, violet!60!black, font=\scriptsize] at (0.80,1.20) {$\eta$};
\draw[very thick, violet!60!black, line cap=round]
  (0.85,1.50) .. controls (1.10,1.47) and (1.30,1.42) .. (1.40,1.52)
  .. controls (1.48,1.57) and (1.55,1.40) .. (1.70,1.40)
  .. controls (1.95,1.37) and (2.15,1.30) .. (2.25,1.50)
  .. controls (2.34,1.65) and (2.42,1.25) .. (2.65,1.25)
  .. controls (3.20,1.20) and (3.80,1.15) .. (4.30,1.10)
  .. controls (4.38,1.02) and (4.46,1.02) .. (4.55,1.10)
  .. controls (5.00,1.10) and (5.50,1.05) .. (5.78,1.02)
  .. controls (5.84,1.15) and (5.88,1.38) .. (5.93,1.10)
  .. controls (6.30,1.00) and (6.80,0.90) .. (7.10,0.85);

\begin{scope}[xshift=6.3cm, yshift=4.55cm]
\node[draw=black, rounded corners=3pt, fill=yellow!10,
      minimum width=2.2cm, minimum height=1.55cm, font=\scriptsize] {};
\node[anchor=north, font=\scriptsize] at (0,0.65) {shared landscape};
\node[anchor=north, font=\scriptsize] at (0,0.35) {$\Jchem(\rho)$};
\draw[blue!70!black, thick]
  (-0.85,-0.25) .. controls (-0.55,-0.20) and (-0.40,-0.05) .. (-0.30,-0.20)
  .. controls (-0.15,-0.30) and (0.00,-0.30) .. (0.10,-0.20)
  .. controls (0.20,-0.10) and (0.30,-0.40) .. (0.45,-0.35)
  .. controls (0.60,-0.30) and (0.75,-0.40) .. (0.85,-0.45);
\draw[->, gray!70, thin] (-0.95,-0.55) -- (0.95,-0.55) node[right,black,font=\tiny] {$\rho$};
\end{scope}

\end{tikzpicture}%
}%
\vfill
\end{tocentry}

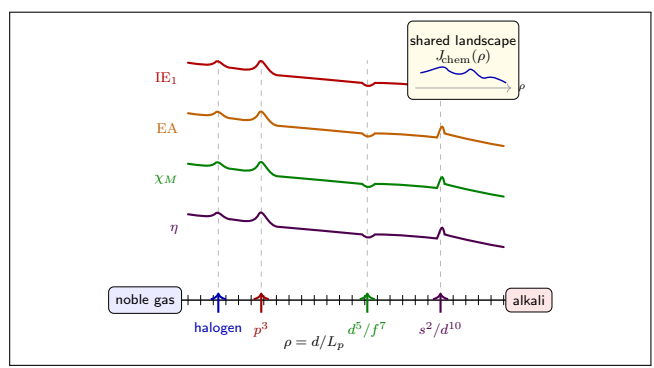
\end{document}